\documentclass[journal]{IEEEtran}
\usepackage{amsmath,amssymb,bm}
\usepackage{array,threeparttable,booktabs,multirow}
\usepackage{graphicx}
\usepackage[ruled]{algorithm2e}
\usepackage[noend]{algpseudocode}
\usepackage{cite}
\usepackage[dvipsnames]{xcolor}
\usepackage{setspace}
\usepackage[colorlinks=true, allcolors=blue]{hyperref}
\usepackage{pifont}
\usepackage{makecell}
\usepackage{diagbox}
\usepackage{setspace}
\usepackage{float}   
\usepackage{subfigure} 
\usepackage{enumitem}

\begin{document} 
\bstctlcite{BSTcontrol}
\linespread{0.88}

\title{
	\huge Deep Generative Model-Aided Power System Dynamic State Estimation and Reconstruction with Unknown Control Inputs or Data Distributions
}
\author{
	Jianhua~Pei,~\IEEEmembership{Student~Member,~IEEE}, %
	Ping Wang,~\IEEEmembership{Fellow,~IEEE}, %
	Jingyu~Wang,~\IEEEmembership{Member,~IEEE}, %
	and~Dongyuan~Shi,~\IEEEmembership{Senior Member,~IEEE} %
}
\maketitle

\begin{abstract}
	Fast and robust dynamic state estimation (DSE) is essential for accurately capturing the internal dynamic processes of power systems, and it serves as the foundation for reliably implementing real-time dynamic modeling, monitoring, and control applications. Nonetheless, on one hand, traditional DSE methods based on Kalman filtering or particle filtering have high accuracy requirements for system parameters, control inputs, phasor measurement unit (PMU) data, and centralized DSE communication. Consequently, these methods often face accuracy bottlenecks when dealing with structural or system process errors, unknown control vectors, PMU anomalies, and communication contingencies. On the other hand, deep learning-aided DSE, while parameter-free, often suffers from generalization issues under unforeseen operating conditions. To address these challenges, this paper proposes an effective approach that leverages deep generative models from AI-generated content (AIGC) to assist DSE. The proposed approach employs an encoder-decoder architecture to estimate unknown control input variables, a robust encoder to mitigate the impact of bad PMU data, and latent diffusion model to address communication issues in centralized DSE. Additionally, a lightweight adaptor is designed to quickly adjust the latent vector distribution. Extensive experimental results on the IEEE 39-bus system and the NPCC 140-bus system demonstrate the effectiveness and superiority of the proposed method in addressing DSE modeling imperfection, measurement uncertainties, communication contingencies, and unknown distribution challenges, while also proving its ability to reduce data storage and communication resource requirements.
\end{abstract}

\begin{IEEEkeywords}
	Generative model, dynamic state estimation, state reconstruction, latent diffusion model, unknown control input, unforeseen distribution, communication contingency.
\end{IEEEkeywords}

\section{Introduction} 
\label{sec:introduction}

\IEEEPARstart{T}{he} stability and reliability of modern power systems are increasingly reliant on accurate and timely dynamic state estimation (DSE) \cite{zhao2020roles}. Specifically, DSE plays a crucial role in providing real-time information about the system's operational status according the updated phasor measurement unit (PMU) data and historical states, which is essential for effective monitoring, control, and decision-making \cite{liu2021dynamic}. The most commonly used DSE approaches are based on traditional filtering techniques for prediction and correction steps\cite{zhao2019power}, such as the extended Kalman filter (EKF) \cite{ghahremani2011dynamic}, unscented Kalman filter (UKF) \cite{qi2016dynamic}, and particle filter (PF) \cite{abolmasoumi2023robust}, where EKF linearizes the system dynamics around the operating point, UKF improves this by better approximating non-linear transformations, and PF addresses non-Gaussian noise and nonlinearities through a probabilistic approach but also introduces additional computational complexity. Nevertheless, these methods rely heavily on accurate system models and well-defined assumptions about system topology and parameters, control inputs, and data distributions \cite{zhao2017robust}. Moreover, in practice, power systems are subject to uncertainties such as unmodeled dynamics, PMU measurement anomalies, cyber-attacks, or communication failures \cite{9946434}, leading to a significant degradation in the performance of conventional DSE approaches \cite{hu2015constrained}.

In response to those DSE challenges, many works have introduced improvements to the aforementioned methods. \cite{zhao2016robust} combines generalized maximum likelihood estimation (GM) and iterative EKF to enable reliable DSE, effectively mitigating large errors, cyber attacks, and temporary disconnections of PMU links during measurement process. Similarly, UKF combined with GM \cite{zhao2017robust}, inequality/equality constraints \cite{zhao2019constrained}, or Koopman operators \cite{netto2018robust}, addresses the issues of non-Gaussian noise distributions and bad PMU data. Some works employ higher-order state estimators \cite{taha2016risk} or vector autoregressive (VAR) \cite{zhao2019correlation} model to jointly predict unknown control vectors, while simultaneously detecting and localizing bad data to mitigate the risks posed by cyber attacks. To address the high computational complexity introduces by PF, event-triggered PF reduces communication bandwidth between decentralized generator nodes and the estimation center \cite{liu2017event, li2018event}. In addition, some DSE approaches focus on applications and enhancements in new electric system scenarios. In \cite{yu2020unscented}, UKF is adapted and deployed for DSE in power systems with wind turbines, while \cite{10477536} designed a variational Bayesian adaptive UKF to estimate the time-varying inertia and damping factors of inverter-based resource-rich power grids.

Compared to model-based DSE, neural networks-enabled data-driven DSE is topology and parameter-free and can fit various data distributions and high nonlinearities, making it a promising research direction. DSE approaches based on artifical neural networks (ANN) \cite{del2007estimation}, long-short-term memory (LSTM) networks \cite{xing2019dynamic}, and variational autoencoder (VAE) \cite{khazeiynasab2021power} can accurately estimate state variables during the transient porcess. However, these learning-based DSE methods inevitably face vulnerabilities to outliers and generalization bottlenecks for unknown data distributions \cite{10345789}. Recently, generative models within the AI-generated content (AIGC), such as generative adversarial networks (GANs) \cite{liu2023spatio}, diffusion models (DMs) \cite{pei2024latent}, and large language models (LLMs) \cite{yan2024hybrid}, have demonstrated powerful capabilities in modeling complex non-linearities and randomness and generating realistic samples. These models are well-suited for handling incomplete, corrupted, or uncertain data \cite{pei2022robust}, making them promising candidates for addressing the limitations of conventional DSE methods and improving power system resilience.

To address the aforementioned challenges in DSE, this paper proposes a novel deep generative model-aided DSE approach that leverages Wasserstein GANs (WGANs), VAEs, and DMs to improve the robustness and accuracy of estimation and reconstruction in power systems. The major contributions are as follows:
\begin{enumerate}[leftmargin=12pt]
	\item A convolutional VAE-WGAN encoder-decoder architecture is developed for state and control variable estimation and data reconstruction in power systems. The VAE-based encoder is a robust encoder for anomalous PMU data, adjusted via projected gradient descent (PGD) convex optimization to mitigate malicious measurement errors.
	\item A latent diffusion model (LDM) installed at the estimation center is used for real-time mitigation of imperfect received low-dimensionallatent vector. The LDM training process adopts a consistent trajectory strategy to ensure one-step or few-step data generation. The sampling process includes two phases: anomaly detection and incomplete data imputation, effectively reducing the communication burden and anomaly impacts.
	\item A lightweight single-layer adaptor is placed between the robust encoder and decoder, allowing learning-based DSE to adapt to unforseen operating conditions through one-shot learning. This enables quick transformations of the latent space based on observations and historical states, improving the generalization performance of DSE under unknown data distributions.
	\end{enumerate}
The extensive experiments on the IEEE 39-bus system and the NPCC 140-bus system demonstrate its superior performance of the proposed DSE approach compared to traditional methods under various challenging scenarios.

The rest of this paper is organized as follows. Problem formulation is presented in Section \ref{sec:SMPF}. Section \ref{sec:DDM} elaborates on the technical implementation of deep generative model-aided DSE. Different aspects of performance demonstrations are given in Section \ref{sec:CS}. Section \ref{sec:Conclusion} concludes the paper. 


\section{System Model and Problem Formulation} 
\label{sec:SMPF}
In this section, DSE state-space  system model is introduced. Subsequently, the existing challenges or open issues in power system DSE are presented in the problem formulation.

\subsection{DSE System Model} \label{DSESM}
The dynamics of the detailed synchronpus generator model consider both a 6th-order GENROU machine and a 2nd-order classical GENCLS model \cite{cui2020hybrid}. In a given power system with $m$ generators, the dynamics of the $i$-th GENROU generator can be characterized by the following differential equations:
\begin{align}
    \frac{d\delta_i}{dt} &= \Delta\omega_i = \omega_i - \omega_0, \label{eq:delta}\\
    \frac{2H_i}{\omega_0}\frac{d\Delta \omega_i}{dt} &= P_{mi} - P_{ei} - D_i\Delta \omega_i/ \omega_0, \label{eq:omega}\\
    T'_{d0i}\frac{dE'_{qi}}{dt} &= -\frac{X_{di} - X''_{di}}{X'_{di} - X''_{di}}E'_{qi} + \frac{X_{di} - X'_{di}}{X'_{di} - X''_{di}}E''_{qi} + E_{fdi}, \label{eq:E_q_prime}\\
    T'_{q0i}\frac{dE'_{di}}{dt} &= -\frac{X_{qi} - X''_{qi}}{X'_{qi} - X''_{qi}}E'_{di} + \frac{X_{qi} - X'_{qi}}{X'_{qi} - X''_{qi}}E''_{di}, \label{eq:E_d_prime}\\
    T''_{d0i}\frac{dE''_{qi}}{dt} &= E'_{qi} - E''_{qi} - (X'_{di} - X''_{di})I_{di}, \label{eq:E_q_d_prime}\\
    T''_{q0i}\frac{dE''_{di}}{dt} &= E'_{di} - E''_{di} - (X'_{qi} - X''_{qi})I_{qi}, \label{eq:E_d_d_prime}
\end{align}
where $\delta_i$ and $\omega_i$ are the $i$-th generator's rotor angle and speed, $\omega_0$ is the rotor speed base value, $H_i$ is the intertia constant, $D_i$ is the damping constant, $P_{mi}$ and $P_{ei}$ are the mechanical power input and the electrical power output (approximately equal to the active power output), $T'_{d0i}$, $T'_{q0i}$, $T''_{d0i}$, and $T''_{q0i}$ are the d/q-axis transient and sub-transient time constants, respectively, in seconds. Moreover, $E'_{di}$, $E'_{qi}$, $E''_{di}$, and $E''_{qi}$ represent the d/q components of internal voltage behind the transient and sub-transient reactance, while $X_{di}$/$X_{qi}$, $X'_{di}$/$X'_{qi}$, and $X''_{di}$/$X''_{qi}$ represent the synchronous, transient, and sub-transient reactances of the generator, respectively. $E_{fdi}$ is the output voltage of the exciter, and $I_{di}$/$I_{qi}$ are the d/q-axis components of the stator current.

\begin{figure}[!h]
	\vspace{-0.2cm}
	\centerline{\includegraphics[width=0.46\textwidth]{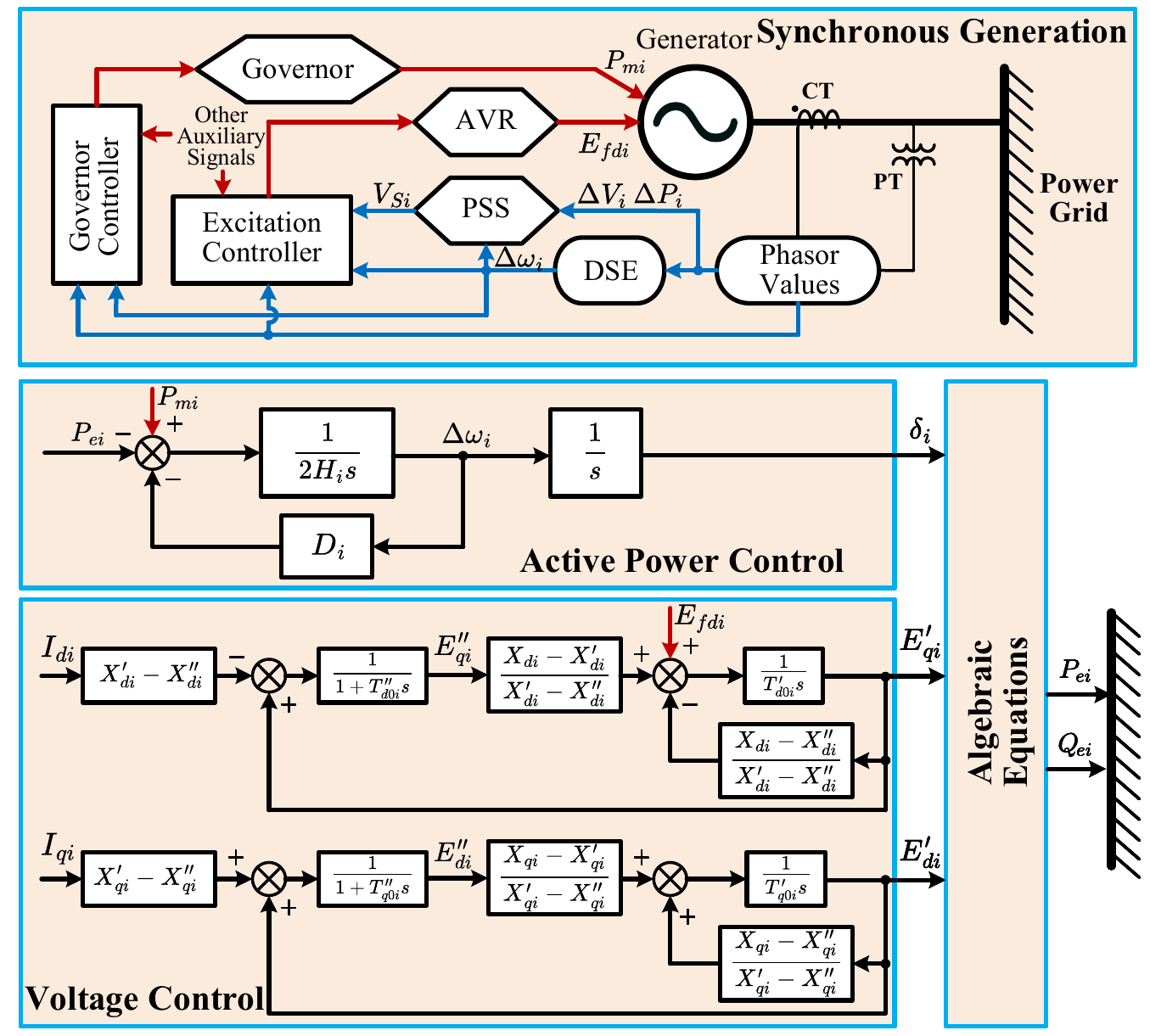}}
	\caption{The position and role of decentralized DSE in power systems, and the composition and control diagram of a 6th-order generator.}
	\label{DSE_fig1}
	\vspace{-0.2cm}
\end{figure}

The algebraic equations are used to model the interconnection of the transmission network with respect to terminal PMU phasor values, which can be detailed as:
\begin{align}
    V_{di} &= V_i\sin(\delta_i-\theta_i),\quad V_{qi} = V_i\cos(\delta_i-\theta_i),\label{eq:Vdq}\\
    V_{di} &= -R_{ai}I_{di} + X''_{qi}I_{qi}+E''_{di},\label{eq:Vd2}\\
    V_{qi} &= -R_{ai}I_{qi} - X''_{di}I_{di}+E''_{qi}, \label{eq:Vq2}\\
    P_{ei} &= V_{di}I_{di} + V_{qi}I_{qi}, \quad Q_{ei} = V_{qi}I_{di} - V_{di}I_{qi}, \label{eq:PQe}\\
    I^2_i &= I^2_{di}+I^2_{qi},\quad \gamma_i = \tan^{-1}(-I_{di}/I_{qi})+\delta_i, \label{eq:Iphasor}
\end{align}
where $V_{di}$/$V_{qi}$ are the d/q components of the generator terminal voltage, $V_i$/$\theta_i$ and $I_i$/$\gamma_i$ are the voltage/current phasors, and $Q_{ei}$ denotes the reactive power output of the generator. The 2nd-order classical GENCLS model simplifies the dynamics by ignoring detailed electromagnetic dynamics and focusing only on mechanical dynamics, as described by Eq. \eqref{eq:delta} and \eqref{eq:omega}. Its algebraic equation can be represented as $P_{ei}=(E_i/X'_{di})\left ( V_i\sin(\delta_i)\cos(\theta_i)-V_i\cos(\delta_i)\sin(\theta_i)\right )$, where $E_ie^{j\delta}$ denotes the voltage behind $X'_{di}$.

Referring to the aforementioned mathematical dynamics of generators, the composition of a single grid-connected generator, i.e., the machine block diagram, is depicted in Fig. \ref{DSE_fig1}. In general, a typical synchronous generator consists of the generator itself, a governor and its controller, and an automatic voltage regulator (AVR) with its exciter controller. The governor outputs $P_{mi}$, controlling the generator's rotor speed and power angle, while the AVR outputs the excitation voltage to control the generator terminal voltage. Additionally, the exciter is typically equipped with a power system stabilizer (PSS) that provides an additional control signal $V_S$ to dampen power system oscillations. Block diagrams for some typical exciters like IEEEX1, governors like TGOV1/TGOV1D, and PSSs like IEEEST can be found in Fig. \ref{DSE_fig2}. After discretizing Eq. \eqref{eq:delta}-\eqref{eq:E_d_d_prime} and those dynamics of the exciters, governors, and stabilizers with a time interval $\Delta t$, the generalized decentralized discrete-time dynamic state-space model of the $i$-th generator is given by
\begin{align}
\bm{x}_{k,i} &= \bm{f}(\bm{x}_{k-1,i}, \bm{y}_{k,i}, \bm{u}_{k,i},  \bm{p}_i)+\bm{w}_{k,i}, \label{eq:stateeq}\\
\bm{z}_{k,i} &= \bm{h}(\bm{x}_{k,i}, , \bm{y}_{k,i}, \bm{p}_i) +\bm{v}_{k,i}, \label{eq:measurementeq}
\end{align}
where $\bm{x}_{k,i}$ represents the state vector,  $\bm{y}_{k,i}$ denotes the algebraic state input vector,  $\bm{u}_{k,i}$ is the control input vector, $\bm{p}_i$ is the system parameters, $\bm{z}_{k,i}$ is the measurement vector typically including voltage/current phasors, bus frequency, and active/reactive power outputs measured by PMUs installed at the generator terminal, $\bm{w}_{k,i}$ and $\bm{v}_{k,i}$ represents the system process and measurement noises assumes to be Gaussian, respectively. Their covariance matrices are denoted by $\bm{Q}_{k,i}$ and $\bm{R}_{k,i}$, respectively. $\bm{f}(\cdot)$ and $\bm{h}(\cdot)$ denote the system process and measurement nonlinear functions. Furthermore, Eq. \eqref{eq:stateeq} represents the state transition equation, which also serves as the prediction step of DSE, while Eq. \eqref{eq:measurementeq} is the observation equation, serving as the filtering and correction step of DSE. As shown in Fig. \ref{DSE_fig1}, DSE is generally deployed in a distributed or centralized manner in the power system to predict and filter-correct the unobservable state variables $\bm{x}_{k,i}$ based on algebraic input and measurement vectors, providing accurate internal state estimates of the generator for the controllers of the exciter, governor, and stabilizer.

\subsection{Problem Formulation}
\label{PF}
\begin{figure}[!h]
	\vspace{-0.2cm}
	\centerline{\includegraphics[width=0.48\textwidth]{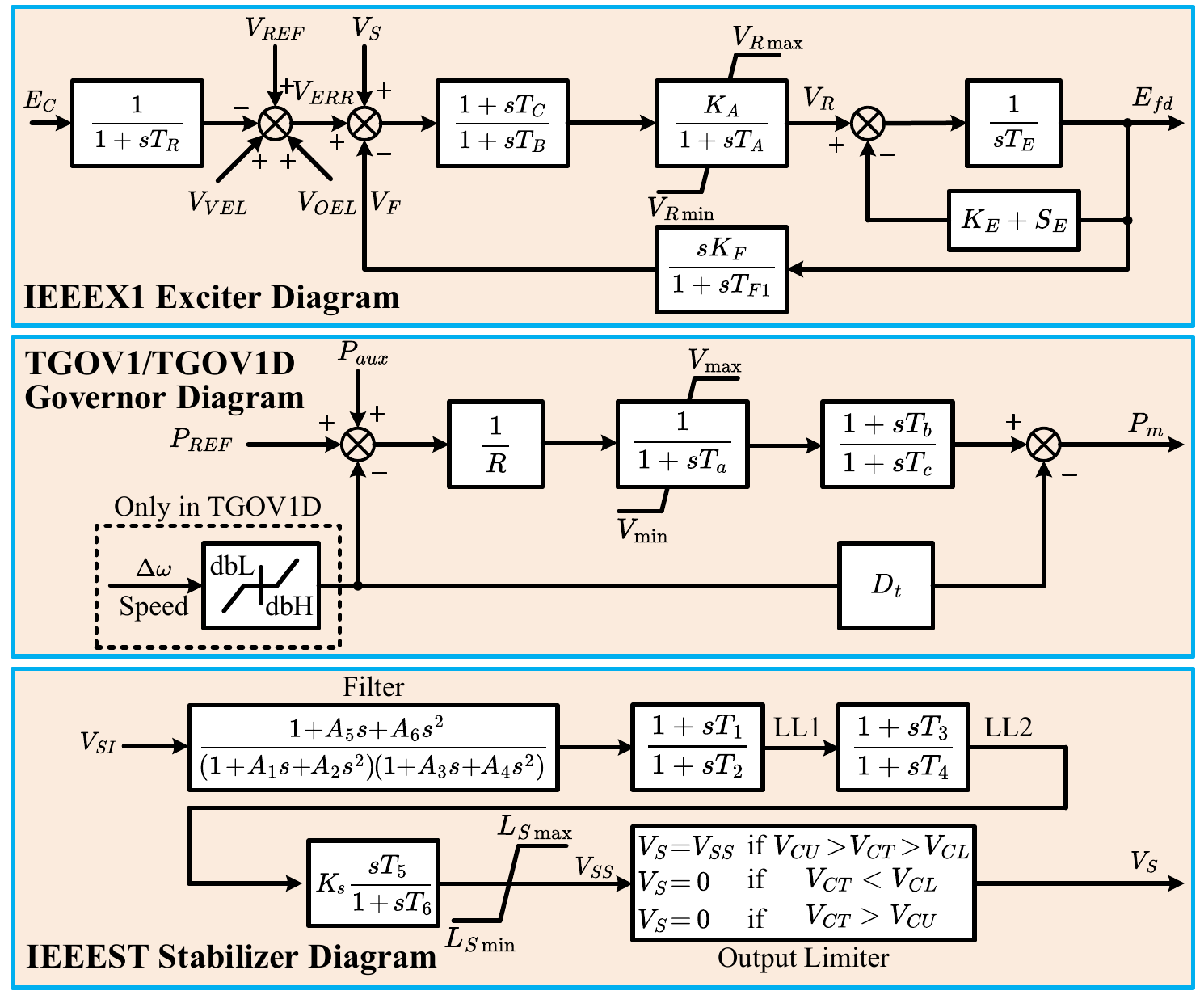}}
	\caption{The control block diagram of the utilized IEEEX1 exciter, TGOV1/TGOV1D governor, and IEEEST stabilizer in this paper, with parameters' definitions described in \cite{cui2020hybrid}.}
	\label{DSE_fig2}
	\vspace{-0.2cm}
\end{figure}

In general, as depicted in the block diagrams in Fig. \ref{DSE_fig2}, different types of exciters, governors, and stabilizers deployed in generators involve highly complex dynamic processes during power system transient events. Considering all the time-discretized equations of these components in Eq. \eqref{eq:stateeq} and \eqref{eq:measurementeq} would undoubtedly increase the computational complexity of the DSE. Moreover, these additional controls involve a large number of system parameters; unknown or inaccurately estimated parameters, as well as dynamically changing parameters, will affect the accuracy of the dynamic state estimation results. Consequently, in some DSE implementations that do not consider the dynamic processes of exciters, governors, and stabilizers, the control inputs $P_{mi}$ and $E_{fdi}$ are assumed to be unknown or not directly measurable. In this case, Eq. \eqref{eq:stateeq} and \eqref{eq:measurementeq} can be rewritten into a nonlinear dynamic state-space model accounts for the control inputs by 
\begin{align}
\bm{x}_{k,i} &= \bm{f}(\bm{x}_{k-1,i}, \bm{y}_{k,i},  \bm{p}_i)+ \bm{G}_i\bm{u}_{k,i}+\bm{w}_{k,i}, \\
\bm{z}_{k,i} &= \bm{h}(\bm{x}_{k,i}, , \bm{y}_{k,i}, \bm{p}_i) +\bm{v}_{k,i}, 
\end{align}
with $\bm{x}_{k,i} = [\delta_{i,k}, \Delta \omega_{i,k}, E'_{qi,k}, E'_{di,k}, E''_{qi,k}, E''_{di,k} ]^{\top}$, $\bm{y}_{k,i} = [ I_{i,k}, \gamma_{i,k} ]^{\top}$, $\bm{u}_{k,i} = [P_{mi,k}, E_{fdi,k}]^{\top}$, 
\begin{align}
\bm{G}_i=\begin{bmatrix}
  0&  0 & \frac{\Delta t}{T'_{d0i}} & 0 & 0 & 0 \\
  0& \frac{\omega_0 \Delta t}{2H_i} & 0 &  0&  0&0
\end{bmatrix}^{\top},
\end{align}
and $\bm{z}_{k,i} = [V_{i,k}, \theta_{i,k}, P_{ei,k}, Q_{ei,k}]^{\top}$. Evidently, the implementation of DSE becomes a challenge with unknown control inputs.

Another challenge for power system DSE comes from outliers commonly present during state prediction and filtering. These outliers can be categorized into observation outliers, innovation outliers, and structural outliers \cite{zhao2016robust}, all of which can significantly impact the effectiveness of the DSE. Observation outliers in $\bm{z}_{k,i}$ may arise from measurement noises that does not follow the assumed Guassian distribution, imperfect PMU measurements, the loss of PMU communication links, and cyber attacks when communicating with the control center or decentralized subsations, such as data loss or delay due to distributed deny of sevice (DDoS) attacks and data manipulation attacks represented by false data injection attacks (FDIA), to name a few. Innovation outliers mainly affect the state prediction process, primarily due to inaccurate or unknown control input vectors or impulsive system process noise, as indicated by Eq. \eqref{eq:stateeq}. Structural outliers are more challenging to handle, caused mainly by system parameter and topology errors, addressing this issue involves joint estimation of model parameters and topology or dual calibration. Handling these three types of outliers simultaneously remains an open issue. Traditional model-based DSE approaches still require further in-depth research to face these challenges, while data-driven DSE methods offer another effective solution to design robust dynamic state estimators in the presence of outliers.

Data-Driven DSEs established based on VAR, LSTM, and ANN break some of the limitations of model-based approaches, such as single-step prediction and specific assumptions about measurement and process noises. Specifically, VAR and LSTM-empowered DSEs utilize a linear or nonlinear function $\bm{\mathcal{F}}(\cdot)$ based on multivariable vectors
\begin{align}
\bm{\mathcal{X}}_{k-p:k-1}&=\begin{bmatrix}
  \bm{x}_{k-p,1} & \cdots  & \bm{x}_{k-1,1} \\
  \vdots & \ddots  & \vdots\\
  \bm{x}_{k-p,m'} & \cdots & \bm{x}_{k-1,m'}
\end{bmatrix} \in \mathbb{R}^{6m' \times p }, 
\end{align}
$\bm{\mathcal{Y}}_{k-p+1:k}\in \mathbb{R}^{4m' \times p}$, $\bm{\mathcal{U}}_{k-p+1:k}\in \mathbb{R}^{2m' \times p}$, and $\bm{\mathcal{Z}}_{k-p+1:k}\in \mathbb{R}^{2m' \times p}$ from past $p$ discrete time steps of $m'$ generators to jointly predict or correct the state vectors $\bm{\mathcal{X}}_{k:k+q-1}$ at the current and future $q$ steps. The estimation results are given by
\begin{equation}
\begin{aligned}
\bm{\mathcal{X}}_{k:k+q-1}  = \bm{\mathcal{F}}(\bm{\mathcal{X}}_{k-p:k-1}, \bm{\mathcal{Y}}_{k-p+1:k}, \bm{\mathcal{U}}_{k-p+1:k}, \bm{\mathcal{Z}}_{k-p+1:k}).
\end{aligned}
\end{equation}
This AI model-driven DSE can be flexibly deployed in a centralized or distributed manner, estimating the state of a single generator or multiple generators. It can perform single-step predictions and filtering or flexibly provide multi-step prediction and calibration for model predictive control (MPC) \cite{10463131}. More importantly, these methods are mostly parameter-free, avoiding structural outliers. Nevertheless, these methods cannot ensure the robustness of state estimation in the presence of observation and innovation outliers. In addition, the interpretability and generalization performance of these learning-based approaches remain a major concern for most researchers. The latter refers to the possibility that machine learning models pre-trained under certain power system conditions may not adapt to unforeseen operating conditions, potentially leading to significant state estimation errors.

In summary, model-based and data-driven DSE in power systems generally still face the following challenges:
\begin{itemize}[leftmargin=10pt]
	\item How can DSE accurately estimate internal state variables and control variables jointly when the generator control inputs $P_{m}$ and $E_{fd}$ are unknown?
	\item How should DSE maintain robustness in state estimation results in the presence of imperfect PMU measurements and communication links, and reduce communication and computational burdens in the case of redundant measurements from multiple generators?
	\item How can learning-based DSE improve the accuracy of state estimation results when facing unforeseen power system operating conditions leading to unknown data distributions and achieve rapid domain adaptation?
\end{itemize}


\section{Deep Generative Model-Aided State Estimation and Reconstruction} 
\label{sec:DDM}
Generative Models can also be flexibly applied to DSE for power systems, effectively addressing issues such as high nonlinearity and weak robustness. However, like other learning-based DSE approaches, deep generative model-assisted DSE also faces generalized challenges. Consequently, strategies for DSE with unknown control inputs are introduced in Subsection \ref{UCI}. Subsection \ref{robustencoder} and Subsection \ref{LDM} apply robust encoders and a latent diffusion model (LDM) to address potential abnormal measurements or communication disturbances. Subsection \ref{Adaptation} designs a method for quick adaptation to unknown operating conditions and Subsection \ref{overallframework} depicts the proposed integral DSE architecture.

\subsection{DSE Approach with Unknown Control Inputs} \label{UCI}
GANs, as an important type of generative model, possess strong capabilities for generating realistic estimated state data. GAN consists of a discriminator $D_{\bm{\gamma}}(\cdot)$ parameterized by $\bm{\gamma}$ and a generator $G_{\bm{\psi}}(\cdot)$ parameterized by $\bm{\psi}$, trained by zero-sum game. The purpose of using GANs is to enable $G_{\bm{\psi}}(\cdot)$ to generate realistic state and observation data $\bm{\mathcal{A}}=[\bm{\mathcal{X}}_{k-p,k+q-1}, \bm{\mathcal{Y}}_{k-p:k+q-1}, \bm{\mathcal{U}}_{k-p:k+q-1}, \bm{\mathcal{Z}}_{k-p:k+q-1}]$, with its adversarial training objective defined as
\begin{equation}
\begin{aligned}
\min_{\bm{\psi}}\max_{\bm{\gamma}}\big \{ & \mathbb{E}_{\bm{\mathcal{A}}\sim q(\bm{\mathcal{A}})}[\log D_{\bm{\gamma}}(\bm{\mathcal{A}})]\\
&+\mathbb{E}_{\textbf{z}\sim q_{\bm{\psi}}(\textbf{z})}[\log(1-D_{\bm{\gamma}}(G_{\bm{\psi}}(\textbf{z})]\big \},
\end{aligned}
\end{equation}
where $q(\bm{\mathcal{A}})$ represents the data distribution of the training historical records of a power system, and $q(\textbf{z}) = \mathcal{N}(\bm{0}, \bm{I})$ denotes the prior distribution of the latent vector $\textbf{z}$. Furthermore, WGAN is established by replacing the Kullback-Leibler $\bm{d}_{KL}$ and Jensen-Shannon divergences $\bm{d}_{JS}$ with the Wasserstein distance $\bm{d}_{W}$ to overcome the challenges of training instability and mode collapse, where its optimization objective can be decoupled into two terms
\begin{equation}
	\begin{aligned}
&\min_{\bm{\psi}} \mathbb{E}_{\textbf{z}\sim q_{\bm{\psi}}(\textbf{z})}\left [ \bm{d}_{W} \left ( \bm{\mathcal{A}} \parallel G_{\bm{\psi}}(\textbf{z}) \right )\right ] \Leftrightarrow - \mathbb{E}_{\textbf{z}\sim q_{\bm{\psi}}(\textbf{z})}\left [ D_{\bm{\gamma}}\left ( G_{\bm{\psi}}(\textbf{z})\right ) \right ]\\
&\min_{\bm{\gamma}}- \mathbb{E}_{\bm{\mathcal{A}} \sim q(\bm{\mathcal{A}})} \left [  D_{\bm{\gamma}}(\bm{\mathcal{A}})\right ] + \mathbb{E}_{\textbf{z}\sim q_{\bm{\psi}}(\textbf{z})}\left [ D_{\bm{\gamma}}\left ( G_{\bm{\psi}}(\textbf{z})\right )\right ].
	\end{aligned}
\end{equation}
In addition, gradient penalty and convolutional layers are considered during the training phase of WGAN, further improving the accuracy of the data synthesized by $G_{\bm{\psi}}(\cdot)$. 

The input to $G_{\bm{\psi}}(\cdot)$ is a low-dimensional latent space $\textbf{z}$. Consequently, to generate deterministic state data, an additional encoder $E_{\bm{\phi}}(\cdot)$ parameterized by $\bm{\phi}$ needs to be trained to determine the latent vector of the current system states. Since the control vector $\bm{\mathcal{U}}_{k:k+q-1}$ is unknown, the input to the convolutional encoder $E_{\bm{\phi}}(\cdot)$ is the matrix $\bm{\mathcal{B}}=[\bm{\mathcal{X}}_{k-p:k-1}, \bm{\mathcal{Y}}_{k-p+1:k}, \bm{\mathcal{U}}_{k-p:k-1}, \bm{\mathcal{Z}}_{k-p+1:k}]$, and the output of $E_{\bm{\phi}}(\cdot)$ can be denoted as $q_{\bm{\phi}}(\textbf{z}|\bm{\mathcal{B}}) \sim \mathcal{N}(\bm{\mu}, \bm{\sigma}^2)$, and $\textbf{z}$ can be reparameterized as $\textbf{z} = \bm{\mu} + \bm{\sigma} \odot \bm{\epsilon}$, where $\bm{\epsilon}  \sim \mathcal{N}(\bm{0},\bm{I})$ and $\odot$ denotes the element-wise product. In this way, as the parameters $\bm{\psi}$ of $G_{\bm{\psi}}(\cdot)$ are kept constant, the evidence lower bound (ELBO) of $E_{\bm{\phi}}(\cdot)$ is 
\begin{equation}
	\begin{aligned}
\mathbb{E}_{q_{\bm{\phi}}(\textbf{z}|\bm{\mathcal{B}})} \left [ \log p_{\bm{\psi}}(\bm{\mathcal{A}}|\textbf{z}) \right ] -\bm{d}_{KL}\left ( q_{\bm{\phi}}(\textbf{z}|\bm{\mathcal{B}})\parallel p_{\bm{\psi}}(\textbf{z}) \right ),
	\end{aligned}
\end{equation}
allowing for encoder parameters update based on the loss
\begin{equation} \label{trainingloss1}
	\begin{aligned}
\mathcal{L}_{\bm{\phi}} = &\alpha_{\bm{\phi}} \underbrace{\mathbb{E} \left \{\bm{d}_{KL}\left [E_{\bm{\phi}}(\bm{\mathcal{B}})\sim \mathcal{N}(\bm{\mu}, \bm{\sigma}^2) \parallel \textbf{z}\sim \mathcal{N}(\bm{0}, \bm{I}) \right ] \right \}}_{\textrm{Prior Fidelity}} \\
 & + (1-\alpha_{\bm{\phi}})\underbrace{\mathbb{E} \left \{\bm{d}_{KL}\left [ G_{\bm{\psi}}(E_{\bm{\phi}}(\bm{\mathcal{B}}) ) \parallel \bm{\mathcal{A}} \right ] \right \}}_{\textrm{Estimation Fidelity}},
	\end{aligned}
\end{equation}
where $\alpha_{\bm{\phi}}$ denotes the encoder loss balance hyper-parameter. This designed encoder-decoder architecture enables the accurate state and unknown control variables estimation by leveraging the complex nonlinear and spatial-temporal correlations both within a single and among multiple generators. Moreover, the deep generative model-assisted DSE only requires the transmission of a low-dimensional $\textbf{z}$ to the substation or control center, which protects data privacy and reduces communication bandwidth.

\subsection{Robust Encoder for Measurement Anomalies} \label{robustencoder}
The uncertainties in phasor data collection and transmission also presents significant challenges for DSE. Specifically, due to the vulnerability of learning-based DSE, the latent space $\textbf{z}$ encoded by $E_{\bm{\phi}}(\cdot)$ based on anomalous measurements $\bm{\mathcal{Z}}$ and phasor values $\bm{\mathcal{Y}}$ in $\bm{\mathcal{B}}$ may be significantly corrupted. Moreover, if $\textbf{z}$ is subjected to data manipulation attacks, signal losses, or delays during communication, it will greatly affect the state and control variables in $\bm{\mathcal{A}}$ decoded by $G_{\bm{\psi}}(\cdot)$. For this reason, a robust encoder $E_{\bm{\phi}'}(\cdot)$ parameterized by $\bm{\phi}'$ at the generator side and additional data processing at the estimation center should be considered.

To design a robust encoder $E_{\bm{\phi}'}(\cdot)$ against contaminated measurements, the vulnerabilities of both the encoder and decoder should be considered simultaneously. The objective for the sufficiently hypothesized small measurement anomalies $\bm{\delta}$ that leads to the maximum estimation and reconstruction error is given by
\begin{equation} \label{objre1}
	\begin{aligned}
\max_{\bm{\delta}} \quad &\bm{d} \left (G_{\bm{\psi}}(E_{\bm{\phi}}(\bm{\mathcal{B}})), G_{\bm{\psi}}(E_{\bm{\phi}}(\bm{\mathcal{B}}+\bm{\delta})) \right )\\
\textrm{s.t.:}  \quad &E_{\bm{\phi}}(\bm{\mathcal{B}}+\bm{\delta}) \sim \mathcal{N}(\bm{0},\bm{I}), \left \|  \bm{\delta} \right \|_p \leq \xi , \bm{L} \leq \bm{\mathcal{B}}+\bm{\delta} \leq \bm{U},
	\end{aligned}
\end{equation}
where $\bm{d}(\cdot, \cdot)$ is the distance function, $\bm{\delta}$ has non-zero values only in updated $\bm{\mathcal{Y}}$ and $\bm{\mathcal{Z}}$ of $\bm{\mathcal{B}}$, $\bm{L}$ and $\bm{U}$ represents the physical upper and lower constraints of the power system, $\left \|  \cdot \right \|_p$ is the p-norm, and $\xi$ denotes the strength of the outliers. When solving Eq. \eqref{objre1}, the objective can be transformed into the standard convex optimization problem
\begin{equation} \label{convexproblem}
	\begin{aligned}
\min_{\bm{\delta}} \quad & \underbrace{\lambda\left \|  \bm{\delta} \right \|_p - \bm{d} \left (G_{\bm{\psi}}(E_{\bm{\phi}}(\bm{\mathcal{B}})), G_{\bm{\psi}}(E_{\bm{\phi}}(\bm{\mathcal{B}}+\bm{\delta})) \right )}_{\bm{e}(\bm{\delta})}\\
\textrm{s.t.:}  \quad &E_{\bm{\phi}}(\bm{\mathcal{B}}+\bm{\delta}) \sim \mathcal{N}(\bm{0},\bm{I}), \bm{L} \leq \bm{\mathcal{B}}+\bm{\delta} \leq \bm{U},
\end{aligned}
\end{equation}
where $\lambda$ is the penalty coefficient. This standard convex optimization problem can be sovled by projected gradient descent (PGD). By leveraging PGD, the values of $\bm{\delta}$ at the $i$-th iteration can be expressed as 
\begin{equation} \label{PGDsolution}
	\begin{aligned}
\bm{\delta}^i = P_C\left (\bm{\delta}^{i-1}- \eta \nabla_{\bm{\delta}} \bm{e}(\bm{\delta}^{i-1})\right ),
	\end{aligned}
\end{equation}
where $P_C(\cdot)$ represents the projection of $\bm{e}(\bm{\delta})$ on the set of constraints $C$. Ultimately, $\bm{\delta}^k$ that causes the worst impact on the encoder and decoder after $k$ iterations and convergence is obtained.

Let $\textbf{z}'\sim \mathcal{N}(\bm{\mu}', \bm{\sigma}'^2)$ be the latent vector encoded based on the contaminated input $\bm{\mathcal{B}}'=\bm{\mathcal{B}}+\bm{\delta}$. According to the smooth encoder in \cite{cemgil2020adversarially} and the training loss \eqref{trainingloss1}, the improved training loss of robust encoder $E_{\bm{\phi}'}(\cdot)$ requires an additional robustness fidelity term to fine-tune the encoder network weights based on obtained $\bm{\delta}$. Hence, the loss function can be rewritten as
\begin{equation} \label{robustloss2}
	\begin{aligned}
\mathcal{L}_{\bm{\phi}'} = &\alpha_{\bm{\phi}} \underbrace{\mathbb{E} \left \{\bm{d}_{KL}\left [\mathcal{N}(\bm{\mu}, \bm{\sigma}^2) \parallel  \mathcal{N}(\bm{0}, \bm{I}) \right ] \right \}}_{\textrm{Prior Fidelity}} \\
 & + (1-\alpha_{\bm{\phi}})\underbrace{\mathbb{E} \left \{  \left \| G_{\bm{\mathcal{\psi}}}(\bm{\mu}+\bm{\sigma}\odot \bm{\epsilon})-\bm{\mathcal{A}} \right \|^2_2  \right \}}_{\textrm{Estimation Fidelity}}\\
& + \underbrace{\mathbb{E}\left \{ \bm{d}_{W,\beta}\left [\mathcal{N}(\bm{\mu}, \bm{\sigma}^2) \parallel  \mathcal{N}(\bm{\mu}', \bm{\sigma}'^2) \right ]\right \}}_{\textrm{Robustness Fidelity}},
	\end{aligned}
\end{equation}
where $\bm{d}_{W,\beta}$ is the the entropy regularized Wasserstein Distance between two distributions and $\beta$ denotes the nonnegative coupling parameter. Specifically, $\bm{d}_{W,\beta}$ can be elucidated as 
\begin{equation} \label{rewdistance}
	\begin{aligned}
&\bm{d}_{W,\beta}\left [\mathcal{N}(\bm{\mu}, \bm{\sigma}^2) \parallel  \mathcal{N}(\bm{\mu}', \bm{\sigma}'^2) \right ] \\
& = \frac{\beta}{2} \left ( \left \| \bm{\mu} -\bm{\mu}' \right \|^2_2 + \bm{\sigma}^2 +\bm{\sigma}'^2 \right )-\frac{1}{2}\log(2\bm{\sigma}^2\bm{\sigma}'^2 ) \\
& \quad + \frac{1}{2}\log \left ( \sqrt{1+4\beta^2\bm{\sigma}^2\bm{\sigma}'^2 } +1 \right) \\
& \quad - \frac{1}{2}\left ( \sqrt{1+4\beta^2\bm{\sigma}^2\bm{\sigma}'^2 } -1\right ) -\log(2\pi)-1.
	\end{aligned}
\end{equation}
As a consequence, the training process for the robust DSE encoder against anomalous PMU measurements can be summarized as Algorithm \ref{trainingrobustencoder}.

\begin{algorithm}[!t]  	\label{trainingrobustencoder}
	\small
	\caption{Training process of DSE robust encoder}
		\LinesNumbered
		\KwIn{Historical operation dataset $q(\bm{\mathcal{A}})$ and $q(\bm{\mathcal{B}})$, learning rate $\eta_1$ and $\eta_2$, encoder $E_{\bm{\phi}}(\cdot)$, decoder $G_{\bm{\psi}}(\cdot)$  }
		\KwOut{ DSE robust fine-tuned encoder $E_{\bm{\phi}'}(\cdot)$ }
		\Repeat{Converged}{
			Sample $\bm{\mathcal{A}} \sim q(\bm{\mathcal{A}})$ and corresponding $\bm{\mathcal{B}} \sim q(\bm{\mathcal{B}})$\;
			Initialize $\bm{\delta}^0 \leftarrow \bm{0}$ and $i \leftarrow 1$\;
			\Repeat{Converged}{
            Compute $\bm{\delta}^i  \leftarrow P_C\left (\bm{\delta}^{i-1}- \eta_1 \nabla_{\bm{\delta}} \bm{e}(\bm{\delta}^{i-1})\right ) $ \;
			Update $i \leftarrow i+1$\;
			}
		Determine $\bm{\delta}$ by $\bm{\delta} \leftarrow \bm{\delta}^k$\;
        Compute $\bm{\mu}, \bm{\sigma}^2 \leftarrow E_{\bm{\phi}'}(\bm{\mathcal{B}})$ and $\bm{\mu}', \bm{\sigma}'^2 \leftarrow E_{\bm{\phi}'}(\bm{\mathcal{B}}+ \bm{\delta})$ \;
        Compute $\bm{d}_{W,\beta}$ by Eq. \eqref{rewdistance} and $\mathcal{L}_{\bm{\phi}'}$ by Eq. \eqref{robustloss2}\;
		Update $\bm{\phi}'$ by $\bm{\phi}' \leftarrow \bm{\phi}'-\eta_2 \nabla_{\bm{\phi}'}\mathcal{L}_{\bm{\phi}'}$\;
		}
\end{algorithm}

\subsection{Latent Space Reconstruction via Latent Diffusion Model} \label{LDM}
The transmitted latent space $\textbf{z}$ may also be corrupted by the imperfect communication links, leading to data losses or blind manipulations in $\textbf{z}$. DMs, due to their powerful generative capabilities and versatility, can be utilized for data enhancement for latent space, and such structures that include an encoder, decoder, and DM are known as LDM. LDM effectively solves the bottleneck of slow sampling speed in DM-based task inference. Concretely, DMs typically consist of a random forward process and a reverse denoising process parameterized by $\bm{\theta}$. The forward process is defined by a unified stochastic differential equation (SDE) 
\begin{equation}
	\begin{aligned}
d\textbf{z} = \bm{u}(\textbf{z},t)dt + \bm{v}(t)d\bm{w}_t,
	\end{aligned}
\end{equation}
where $\bm{u}(\textbf{z},t)$ and $\bm{v}(t)$ are the drift and diffusion coefficients, and $\bm{w}_t$ represents a standard Wiener process. These coefficients can be designed differently for variance-preserving (VP) and variance-exploding (VE) formulations. By considering the marginal distribution, the probability flow ordinary differential equation (ODE) is defined as
\begin{equation} \label{ODE}
	\begin{aligned}
d\textbf{z} = \left [\bm{u}(\textbf{z},t)-\frac{1}{2} \bm{v}^2(t) \nabla_{\textbf{z}}\log p_t(\textbf{z}_t) \right ]dt,
	\end{aligned}
\end{equation}
where $\nabla_{\textbf{z}}\log p_t(\textbf{z}_t)$ denotes the score function. In \cite{karras2022elucidating}, Eq. \eqref{ODE} can be expressed in a more general form as
\begin{equation} \label{EDMform}
	\begin{aligned}
d\textbf{z} = \left [  \frac{\dot{\bm{s}}(t)}{\bm{s}(t)}\textbf{z} - \bm{s}(t)^2\dot{\bm{\sigma}}(t)\bm{\sigma}(t)\nabla_{\textbf{z}}\log p_t\left( \frac{\textbf{z}_t}{\bm{s}(t)}, \bm{\sigma}(t)\right ) \right ]dt,
	\end{aligned}
\end{equation}
where $\bm{s}(t)$ is an additional scale schedule, and $\bm{\sigma}(t)$ defines the noise level at time $t$. Specifically, according to elucidating DM (EDM) \cite{karras2022elucidating}, $\bm{s}(t)=1$, $\bm{\sigma}(t)=t$, and the discrete $N$-step noise schedule is given by $\left \{ \sigma_n \right \}^{n=N}_{n=0}$, where $\sigma_n=0$ when $n=0$, and $\sigma_n = \left(\sigma_1^{1/\rho} + \frac{n-1}{N-1}(\sigma_N^{1/\rho}- \sigma_1^{1/\rho})  \right )^{\rho}$ when $n>0$. Here, $\sigma_1=0.002$, $\sigma_N=80$, and $\rho=7$. Evidently, the forward process of DM can be written as $q(\textbf{z}_t|\textbf{z}_0) = \mathcal{N}(\textbf{z}_t; \textbf{z}_0, t^2\bm{I})$ according to Eq. \eqref{EDMform}, and this process can also be simplified as the sampling form $\textbf{z}_t = \textbf{z}_0+ \bm{n}$, where $\bm{n}\sim \mathcal{N}(\bm{0}, t^2\bm{I})$.

In the reverse process, the score function $\nabla_{\textbf{z}}\log p_t(\textbf{z}_t)$ can be approximated by $\left (D_{\bm{\theta}}(\textbf{z}_t, t)- \textbf{z}_t\right ) /t^2$, where $D_{\bm{\theta}}(\textbf{z}_t, t)$ is the denoiser function. As a result, the reverse process can be simplified as 
\begin{equation}
	\begin{aligned}
d\textbf{z} = - \left (D_{\bm{\theta}}(\textbf{z}_t, t)- \textbf{z}_t\right ) /t dt.
	\end{aligned}
\end{equation}
Specifically, $D_{\bm{\theta}}(\textbf{z}_t, t)$ can be obtained by minimizing $L_2$ distance between denoising output and clean data, which is given by $\mathbb{E}_{\bm{\epsilon}\sim \mathcal{N}(\bm{0}, \bm{I}), n \sim \mathcal{U}[1,N]}\left [ \lambda (t_n)\left \| D_{\bm{\theta}}(\textbf{z}+ t_n\bm{\epsilon}, t_n) - \textbf{z} \right \|^2_2 \right ]$, where $\lambda (t_n)$ is the loss weight. Nonetheless, this direct matching method often results in poor generated data, so many works choose to train a neural network $F_{\bm{\theta}}(\textbf{z}_t, t)$ to build $D_{\bm{\theta}}$, which is given by 
\begin{equation} \label{skipconnection}
	\begin{aligned}
D_{\bm{\theta}}(\textbf{z}_t, t) = c_{\textrm{skip}}(t)\textbf{z}_t+c_{\textrm{out}}(t)F_{\bm{\theta}}(c_{\textrm{in}}(t)\textbf{z}_t, c_{\textrm{noise}}(t)),
	\end{aligned}
\end{equation}
where $c_{\textrm{skip}}(t)$ modulates the skip connection, $c_{\textrm{in}}(t)$ and $c_{\textrm{out}}(t)$ scale the input and output magnitudes, and $c_{\textrm{noise}}(t)$ maps the noise level $t$ into a conditioning input for $F_{\bm{\theta}}$. The emergence of the consistency model has greatly overcome the time-consuming bottleneck of multi-step predictions in diffusion models, as it can directly map $\textbf{z}_{t_n}(n>1)$ to $\textbf{z}_{\varepsilon}$ to achieve high-quality data generation in one or a few steps, where $\varepsilon=t_1$. The main contribution of the proposed approach is the combination of consistency model training loss and decoder output performance, and the training loss is rewritten as 
\begin{equation} 
	\begin{aligned}
&\mathcal{L}_{\bm{\theta}}\left( \bm{\theta}, \bm{\theta}^{-}| \bm{\psi}\right )=\\
&\mathbb{E}_q \left [ \left \|G_{\bm{\psi}}\left (D_{\bm{\theta}}(\textbf{z}+t_{n+1}\bm{\epsilon}, t_{n+1}) \right )- G_{\bm{\psi}}\left (D_{\bm{\theta}^-}(\textbf{z}+t_{n}\bm{\epsilon}, t_{n}) \right ) \right \|^2_2\right ],
	\end{aligned}
\end{equation}
where $\bm{\theta}^-$ denotes a runing average of the past values of $\bm{\theta}$. Clearly, the consisteny model's denoiser $D_{\bm{\theta}}(\textbf{z}_t, t)$ trains the network by mapping adjacent noisy data points $t_{n+1}$ and $t_n$ along the same forward diffusion trajectory to the same $\textbf{z}_{\varepsilon}$, ultimately obtaining a single-step generative model.

\begin{figure}[!h]
	\vspace{-0.2cm}
	\centerline{\includegraphics[width=0.48\textwidth]{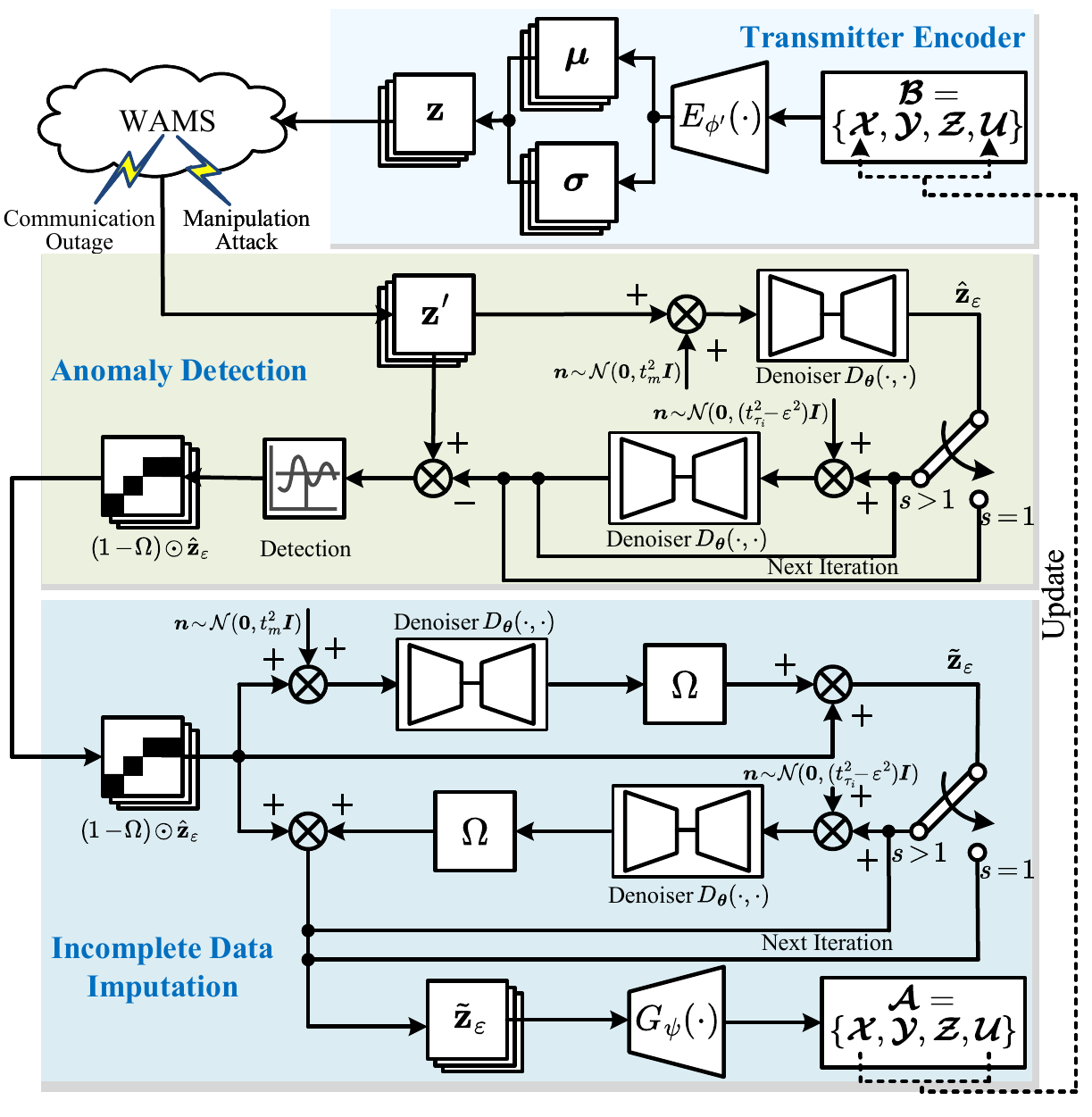}}
	\caption{The enhanced two-phase latent diffusion model for dynamic state estimation, composed of a robust encoder $E_{\bm{\phi}'}(\cdot)$, a TSDM model trained in the latent space $D_{\bm{\theta}}(\cdot, \cdot)$, and the generator $G_{\bm{\psi}}(\cdot)$ of GAN, which aims to mitigate the impact of communication uncertainty in centralized DSE.}
	\label{DSE_fig3}
	\vspace{-0.2cm}
\end{figure}

Unlike generation tasks, the task of DSE is to improve the quality of the reconstructed data. Fortunately, in addition to performing well in data generation, the consistency model also supports zero-shot data editing, such as image inpainting and super-resolution. Therefore, for potential data losses and manipulations, the proposed DSE strategy, as shown in Algorithm \ref{communicationenhancement} and Fig. \ref{DSE_fig3}, is a two-phase mitigation method using $D_{\bm{\theta}}(\cdot, \cdot)$ for anomaly detection and incomplete data imputation. In the detection phase, the received raw encodings $\textbf{z}'$ first undergo a forward process to become the noisy data $\textbf{z}_{t_m}=\textbf{z}'+t_m\bm{\epsilon}$ based on the selected noise level $t_m$. Subsequently, $\textbf{z}_{t_m}$ is transformed into $\hat{\textbf{z}}_{\varepsilon}$ through one or more steps of the reverse process based on the subsampling sequence $\bm{\tau}=[\tau_1, \tau_2, \cdots, \tau_s]$. Thus, anomalies can be detected by observing $\textbf{z}'-\hat{\textbf{z}}_{\varepsilon}$. If the error value exceeds the detection threshold, the received data in those spaitial-temporal positions will be discarded. The locations of discarded data are denoted by $\Omega$. In the imputation phase, the missing entries are first filled with zeros, then undergo a forward process with noise of variance $t^2_m$. The noisy data is mapped to $\tilde{\textbf{z}}_{\varepsilon}$ via $D_{\bm{\theta}}(\cdot, \cdot)$. The generated data $\tilde{\textbf{z}}_{\varepsilon}$ is combined with the correct observations at the next subsampling point by $\hat{\textbf{z}}_{\varepsilon} \odot (1-\Omega)  +  \tilde{\textbf{z}}_{\varepsilon}\odot \Omega$. Notably, the proposed DSE communication enhancement method only utlize 0 to $m$ portion of the pretrained consistency model, allowing the original schedule to be truncated during training to save computational resources.

\begin{algorithm}[!t]  	\label{communicationenhancement}
	\small
	\caption{DSE Communication Enhancement}
		\LinesNumbered
		\KwIn{ Consistency denoiser $D_{\bm{\theta}}(\cdot, \cdot)$, subsequence length $s$ and time points $\tau_1=m>\tau_2>\cdots>\tau_s$, transmitted data $\bm{\mathcal{B}}$, robust encoder $E_{\bm{\phi}'}(\cdot)$, decoder $G_{\bm{\psi}}(\cdot)$, detection threshold $\mathcal{T}$}
		\KwOut{Reconstructed and estimated $\bm{\mathcal{A}}$ in estimation center }
		Compute the encoded latent space $\textbf{z} \leftarrow E_{\bm{\phi}'}(\bm{\mathcal{B}})$\;
		Transmit $\textbf{z}$ over the corrupted communication links and contaminated $\textbf{z}'$ is received\;
        Sample noisy data by $\hat{\textbf{z}}_{t_m} \sim \mathcal{N}\left (\hat{\textbf{z}}_{t_m}; \textbf{z}', t_m^2\bm{I}   \right)$ \;
        Compute denoised estimation $\hat{\textbf{z}}_{\varepsilon} \leftarrow D_{\bm{\theta}}(\hat{\textbf{z}}_{t_m}, t_m)$\;
		\If{$s>1$}{
			\For{$i = 2$ \textup{\textbf{to}} $s$}{
                Sample $\hat{\textbf{z}}_{t_{\tau_i}} \sim \mathcal{N} (\hat{\textbf{z}}_{t_{\tau_i}}; \hat{\textbf{z}}_{\varepsilon}, (t_{\tau_i}^2-\varepsilon^2 )\bm{I})$\;
				Compute $\hat{\textbf{z}}_{\varepsilon} \leftarrow D_{\bm{\theta}}(\hat{\textbf{z}}_{t_{\tau_i}}, t_{\tau_i})$\;
			}
		}
        Determine corrupted positions $\Omega$ by $\left | \textbf{z}'- \hat{\textbf{z}}_{\varepsilon}\right | >\mathcal{T}$\;
        Combine by $\tilde{\textbf{z}} \leftarrow \hat{\textbf{z}}_{\varepsilon} \odot (1-\Omega) + \bm{0}\odot \Omega $\;
        Sample noisy data by $\tilde{\textbf{z}}_{t_m} \sim \mathcal{N}\left (\tilde{\textbf{z}}_{t_m}; \tilde{\textbf{z}}, t_m^2\bm{I}   \right)$ \;
        Compute denoised estimation $\tilde{\textbf{z}}_{\varepsilon} \leftarrow D_{\bm{\theta}}(\tilde{\textbf{z}}_{t_m}, t_m)$\;
        Combine by $\tilde{\textbf{z}}_{\varepsilon} \leftarrow \hat{\textbf{z}}_{\varepsilon} \odot (1-\Omega) + \tilde{\textbf{z}}_{\varepsilon}\odot \Omega $\;
        \If{$s>1$}{
			\For{$i = 2$ \textup{\textbf{to}} $s$}{
                Sample $\tilde{\textbf{z}}_{t_{\tau_i}} \sim \mathcal{N} (\tilde{\textbf{z}}_{t_{\tau_i}}; \tilde{\textbf{z}}_{\varepsilon}, (t_{\tau_i}^2-\varepsilon^2 )\bm{I})$\;
				Compute $\tilde{\textbf{z}}_{\varepsilon} \leftarrow D_{\bm{\theta}}(\tilde{\textbf{z}}_{t_{\tau_i}}, t_{\tau_i})$\;
                Combine by $\tilde{\textbf{z}}_{\varepsilon} \leftarrow \hat{\textbf{z}}_{\varepsilon} \odot (1-\Omega) + \tilde{\textbf{z}}_{\varepsilon}\odot \Omega $\;
			}
		}
		\textbf{Return} the DSE output by $\bm{\mathcal{A}} \leftarrow G_{\bm{\psi}}(\tilde{\textbf{z}}_{\varepsilon})$
\end{algorithm}

\subsection{Fast Adaptation for Unaware Power System Events} \label{Adaptation}
\begin{figure*}[!t]
\vspace{-0.2cm}
\centerline{\includegraphics[width=0.85\textwidth]{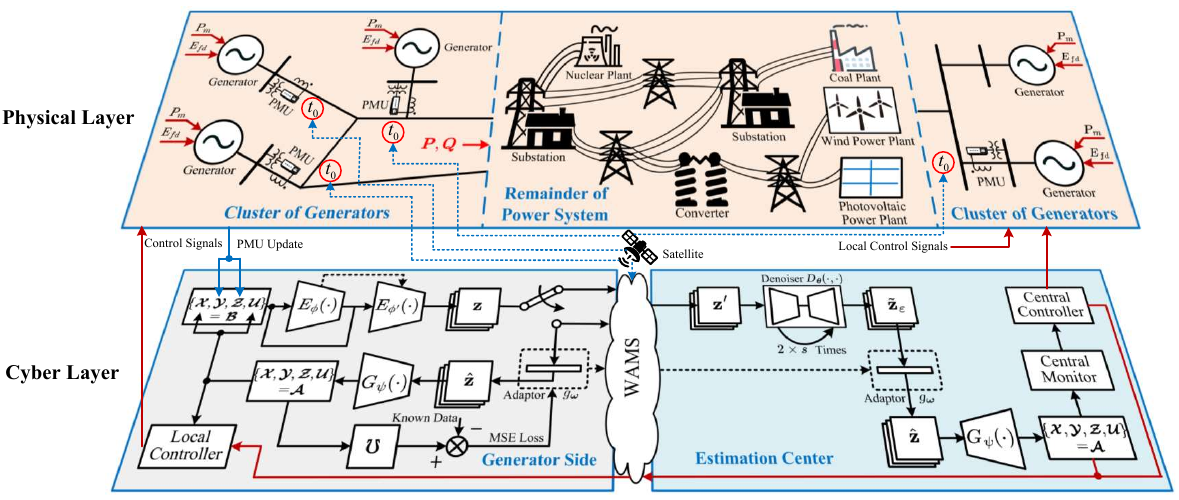}}
\caption{The deployment diagram of the deep generative model-aided data-driven DSE, which integrates both centralized and decentralized approaches in practical power systems. This can be used for local, accurate, and robust DSE on the power plant side, as well as monitoring in estimation centers, ensuring the stable and secure operation of both local and global control systems.}
\label{DSE_fig4}
\vspace{-0.5cm}
\end{figure*}

Learning-based DSE algorithms will degrade in estimation accuracy under unforeseen operating conditions. Based on the architecture of LDM, a lightweight single-layer neural network called an adaptor can be placed after the encoder to transform the encoded latent space, adapting to new operating conditions of the power system. Specifically, adaptor $\bm{g}_{\bm{\omega}}(\cdot)$ is parameterizd by $\bm{\omega}$ and placed between the encoder and the decoder as illustrated in Fig. \ref{DSE_fig4}. As previously discussed, some data in $\bm{\mathcal{A}}$ generated by $G_{\bm{\psi}}(\cdot)$ is known, such as algebraic input $\bm{\mathcal{Y}}$ and measurement vectors $\bm{\mathcal{Z}}$, the known data positions in $\bm{\mathcal{A}}$ are denoted as $\mho$. Decoder $G_{\bm{\psi}}(\cdot)$ is deployed at both the generator node and the estimation center. If the decoding error of $\bm{\mathcal{A}} \odot \mho$ exceeds the threshold, it indicates that DSE has encountered an unknown data distribution, and $\bm{g}_{\bm{\omega}}(\cdot)$ can be activated to perform one-shot or few-shot learning, transforming $\textbf{z}$ by 
\begin{equation} 
	\begin{aligned}
\hat{\textbf{z}} = \bm{g}_{\bm{\omega}}(\textbf{z}) = \bm{\omega}^{\top} \textbf{z} + \bm{b},
	\end{aligned}
\end{equation}
where $\bm{b}$ denotes the bias of the adaptor. Moreover, the loss for training $\bm{g}_{\bm{\omega}}(\cdot)$ only considers the reconstruction error in $\mho$, which is given by
\begin{equation} 
	\begin{aligned}
\mathcal{L}_{\bm{\omega}} = \mathbb{E}_q \left [  \left \| G_{\bm{\psi}} \left ( \bm{g}_{\bm{\omega}} \left ( E_{\bm{\phi}'}(\bm{\mathcal{B}} ) \right ) \right )  \odot \mho - \bm{\mathcal{A}}\odot \mho \right \|^2_2 \right ]
	\end{aligned}
\end{equation}

In the proposed DSE framework, when estimating the state under known operating conditions, $\bm{g}_{\bm{\omega}}(\cdot)$ will be removed, and $\textbf{z}$ will be sent directly to the receiving end. When rapid domain adapatation is activated, the original $\textbf{z}$ and adaptor parameters $\bm{\omega}$ will be sent together to the estimation center to avoid changing the parameters of robust encoder $\bm{\phi}'$, decoder $\bm{\psi}$, and denoiser $\bm{\theta}$. Consequently, $\bm{g}_{\bm{\omega}}(\cdot)$ is defined as a dynamic lightweight single-layer neural network to complete the latent space transformation. Due to the low dimensionality of the latent space and the given initial encoding $\textbf{z}$, this domain adaptation introduces only minimal computational complexity and communication bandwidth burden.

\subsection{Cyber-Physical DSE Implementation} \label{overallframework}
The proposed DSE framework is capable of estimating unknown control vectors, mitigating the impact of PMU measurement and communication anomalies, and improving state estimation accuracy under unforeseen operating conditions. As depicted in Fig. \ref{DSE_fig4}, PMUs are installed at the generator terminal buses, and their measurements update the measurement matrix $\bm{\mathcal{Z}}$ and the algebraic input matrix $\bm{\mathcal{Y}}$. Assuming that a cluster of generators in a certain region has $m'$ units, input data $\bm{\mathcal{B}}$ will be encoded by a robust encoder into a low-dimensional latent space $\textbf{z}$, where the robust encoder is resilient to anomalous PMU data inputs. $\textbf{z}$ is transmitted from the generator nodes to the estimation center via the wide area measurement system (WAMS). 

The denoiser $D_{\bm{\theta}}(\cdot, \cdot)$ at estimation center performs $2\times s$ rounds of denoising processes to restore the received $\textbf{z}'$. When the power system is in an unknown operating state, real-time adaptive learning of $\bm{g}_{\bm{\omega}}(\cdot)$ will be triggered to update the parameters $\bm{\omega}$, which will be transmitted along with $\textbf{z}$ to the estimation center for transformation via a dynamic neural network. Utimately, the decoder $G_{\bm{\phi}}(\cdot)$ will generate $\bm{\mathcal{A}}$, which includes the state values $\bm{x}_k$ and control vectors $\bm{u}_k$ at time $k$ with reconstructed $\hat{\textbf{z}}$, enabling monitoring of the generator cluster’s states and issuing global control signals to local controllers.

\section{Experimental Results}  
\label{sec:CS}
\subsection{Experimental Setup}
\textbf{Data Source:} The Experimental data are obtained from IEEE 10-machine 39-bus test system and NPCC 48-machine 140-bus test system, both established on the open-source Python-based power system transient simulation platform ANDES \cite{cui2020hybrid}. Specifically, in IEEE 39-bus system, all 10 generators are GENROU type, equipped with IEEEX1 exciters, IEEEST stabilizers, and TGOV1 governors. The NPCC 140-bus test system includes 21 GENCLS generators and 27 GENROU generators, equipped with IEEEX1 exciters and TGOV1 governors. A total of 5000 diverse power system events, including short-circuit faults, line trips, generator sheddings, and load changes, were simulated in two systems. The occurrence times for these events were uniformly set at 1s, while other parameters such as location, duration, grounding impedance, number of cut-off and reclosed lines, and load changes were randomly selected within reasonable ranges.

\textbf{Baseline Methods:} To demonstrate the superiority of the proposed DSE approach, it should be compared with existing methods. For DSE accuracy, two traditional model-based DSE approaches are implemented: the conventional UKF and an improved UKF enabled by VAR for control variable estimation (UKF-VAR) \cite{zhao2019correlation}. Additionally, machine learning-assisted DSE methods represented by VAE-based approaches \cite{khazeiynasab2021power}, are considered to validate the advancements of the proposed method. In terms of real-time performance and data compression for DSE, a non-latent space and multi-step two stage diffusion model (TSDM)-based data recovery method \cite{10542391} was implemented to demonstrate the low computational complexity and communication bandwidth of the proposed DSE approach.

\textbf{Model Training:} All neural network training was conducted using Python 3.8.19 and CUDA-accelerated PyTorch 2.3.0 on a computer equipped with an i5-13600KF CPU operating at 3.50GHz, 32 GB of RAM, and an NVIDIA GeForce RTX 4070 GPU. The training steps for the proposed DSE model include training the WGAN, training the VAE as the inversion of WGAN, optimizing the robust encoder, and training the LDM. Notably, to reduce training costs, the schedule length $m$ of the LDM is $m=50$. Additionally, the training set consists of 3000 samples involving short-circuit faults, line trips, and load changes, while the test set comprises 1250 samples of generator sheddings and the remaining samples of other events.

\begin{figure*}[!h]
	\vspace{-0.2cm}
	\centerline{\includegraphics[width=0.7\textwidth]{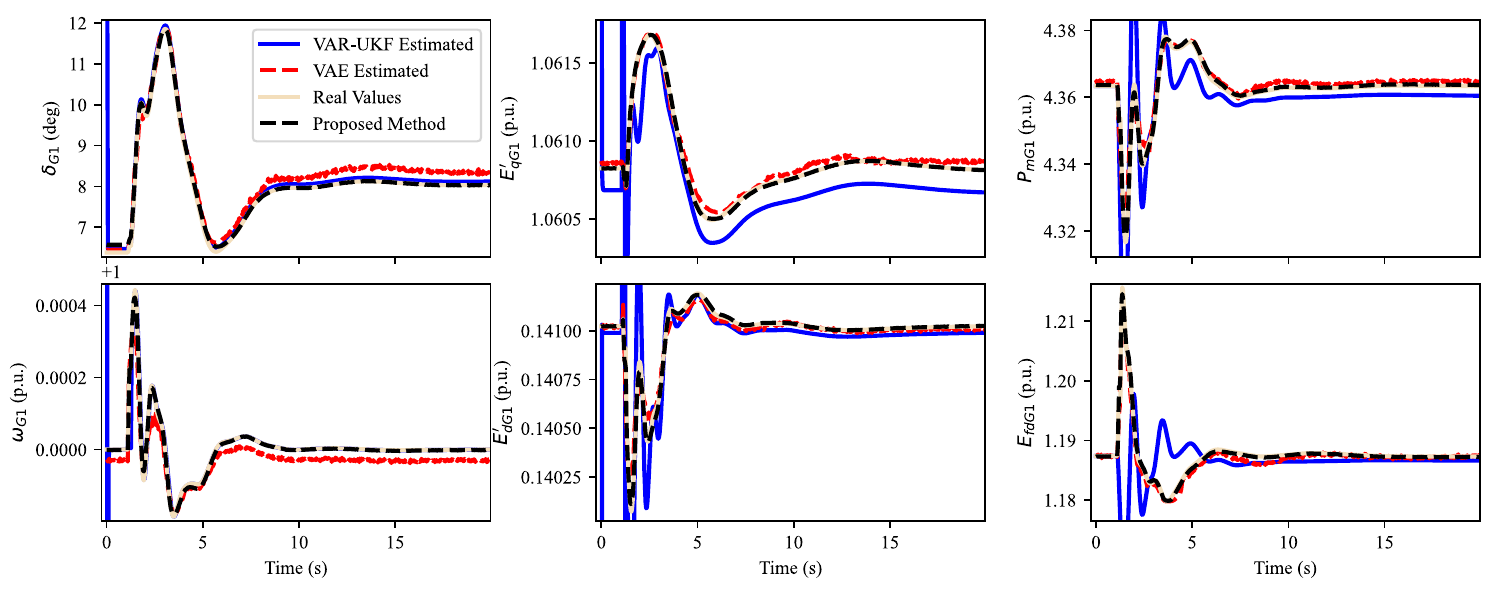}}
	\caption{Joint state and control variable estimation results for Generator 1 during a short-circuit fault event in the IEEE 39-bus system.}
	\label{res_fig1}
	\vspace{-0.2cm}
\end{figure*}

\begin{figure*}[!h]
	\vspace{-0.2cm}
	\centerline{\includegraphics[width=0.7\textwidth]{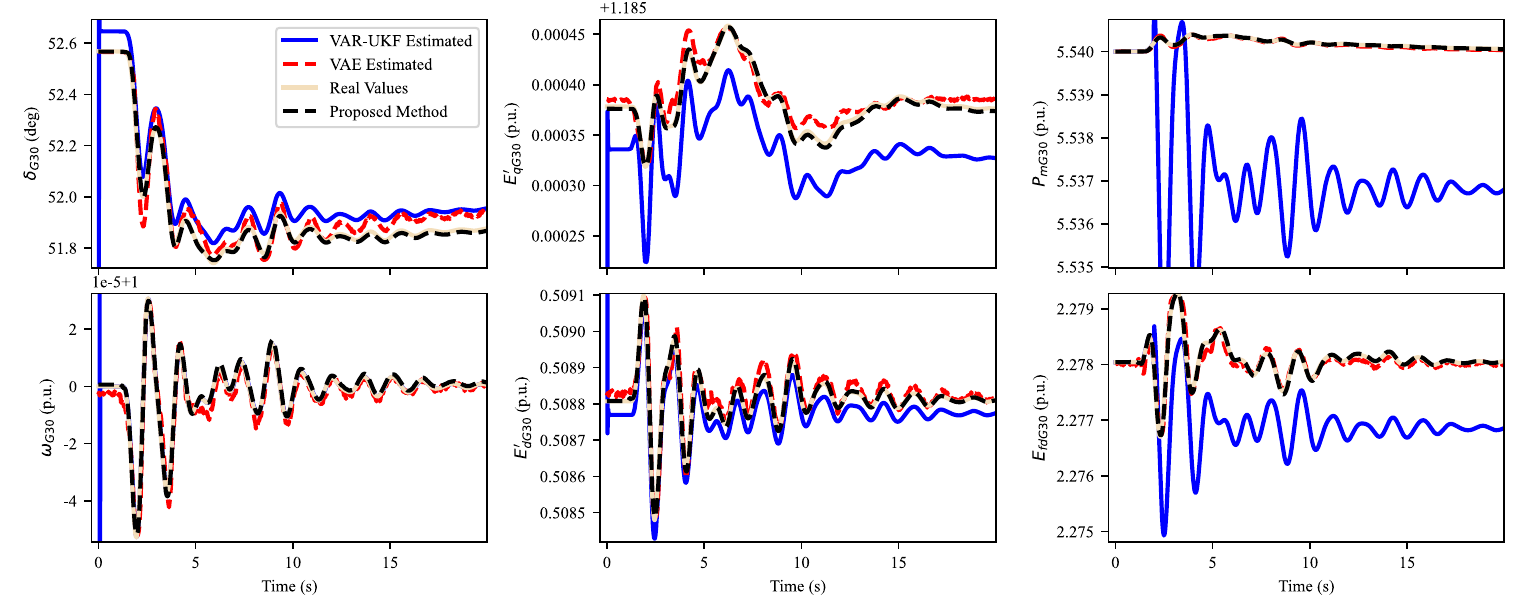}}
	\caption{Joint state and control variable estimation results for Generator 30 during a load disturbance event in the NPCC 140-bus system..}
	\label{res_fig2}
	\vspace{-0.5cm}
\end{figure*}

\subsection{Estimation Accuracy with Unknown Control Inputs}
As shown in Fig. \ref{DSE_fig2}, the control outputs of the generator exciter, governor, and stabilizer, which require a large number of accurate parameters, may be inaccurate or unknown. The proposed approach is tested on both the IEEE 39-bus system and the NPCC 140-bus system, using a comparison between DSE based on UKF-VAR and VAE. In the IEEE 39-bus system, a single three-phase fault occurs at Bus 22 with a fault duration of 0.07s and a fault impedance of $x_f$ = 0.09 p.u.. The actual and estimated vaules of the internal state variables ($\delta, \omega, E'_q, E'_d$) and control variables ($P_m, E_{fd}$) for Generator 1 (G1, terminal Bus 30) are depicted in Fig. \ref{res_fig1}, where both $\bm{Q}$ and $\bm{R}$ are set to $1e^{-4}\bm{I}$. It is evident that the traditional VAR-UKF method requires convergence to the estimated values at the initial time, and its estimation of the power angle and rotor speed has similar accuracy to the proposed method, while the VAE-based DSE performs slightly worse. However, in the estimation of control variables and transient voltage, the traditional UKF method shows a noticeable deviation, primarily due to inaccurate input controls. VAR struggles with the strong nonlinear prediction challenge, whereas the VAE and the proposed method, which use the same estimation strategy, are more effective at addressing this bottleneck.

Similar tests are also conducted on the NPCC 140-bus system, with the difference being that at 1s, a sudden loss of 0.067 p.u. of active load occurs at Bus 19. The estimation results for the state variables and control variables of Generator 30 (G30, terminal Bus 36) are illustrated in Fig \ref{res_fig2}. In this test case, since the variations in the generator's state and control variables are small, the estimation errors of VAR-UKF are relatively large. On the other hand, the proposed method achieves a higher accuracy in jointly estimating the strongly nonlinear state and control variables.

\begin{figure*}[!t]
    \centering
    \subfigure[Estimated states with PMU data containing ramp error components during a line trip event in the IEEE 39-bus system.]{
        \includegraphics[width=0.315\textwidth]{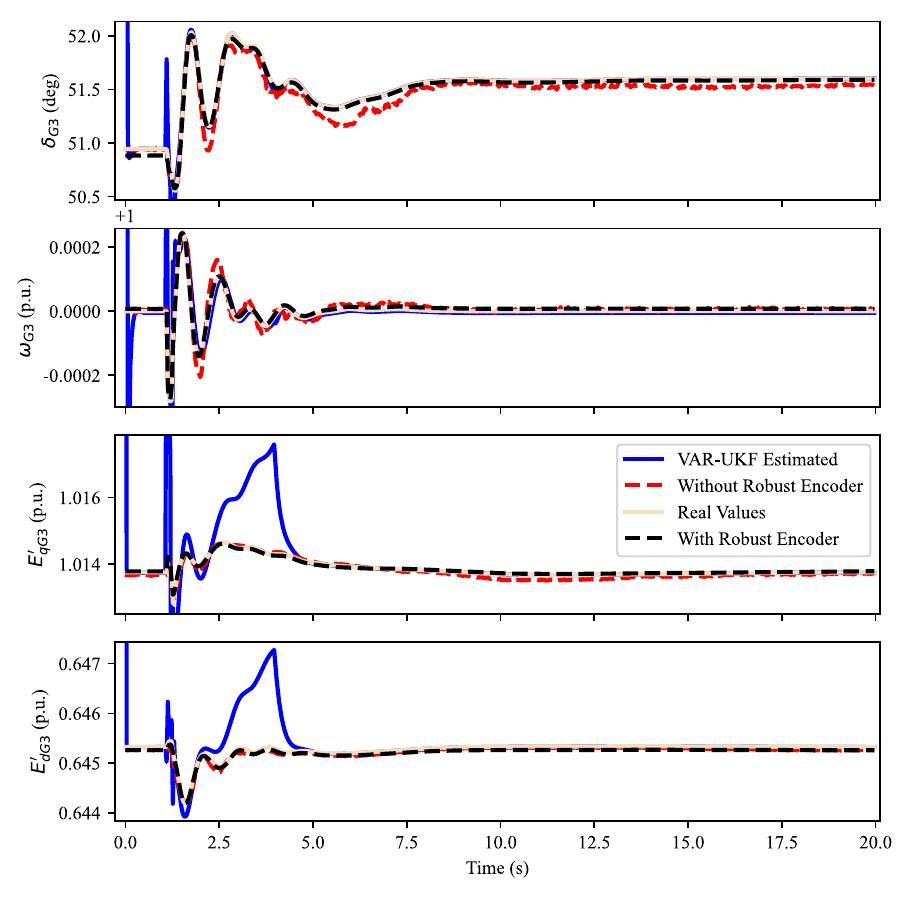}
        \label{res_fig3}
    }
    \subfigure[Estimated states with PMU data containing step error components during a short-circuit fault and line trip event in the NPCC 140-bus system.]{
        \includegraphics[width=0.315\textwidth]{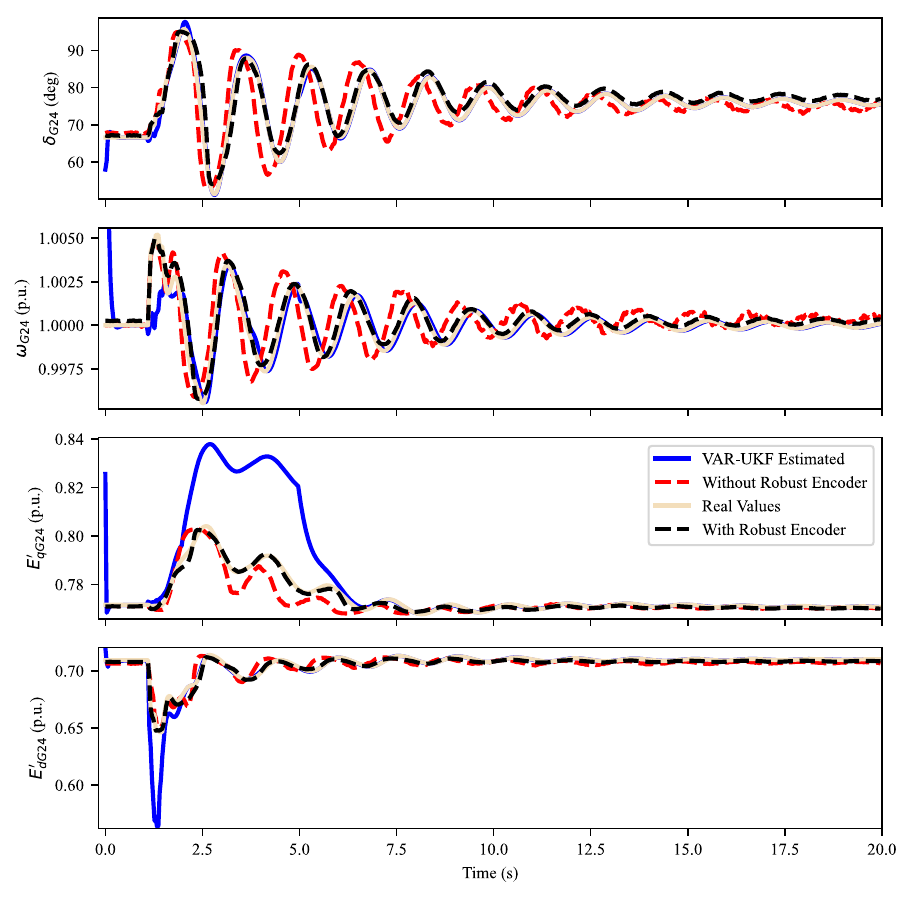}
        \label{res_fig5}
    }
    \subfigure[Estimated states with 50\% of transmitted data contaminated during a load disturbance event in the IEEE 39-bus system.]{
        \includegraphics[width=0.315\textwidth]{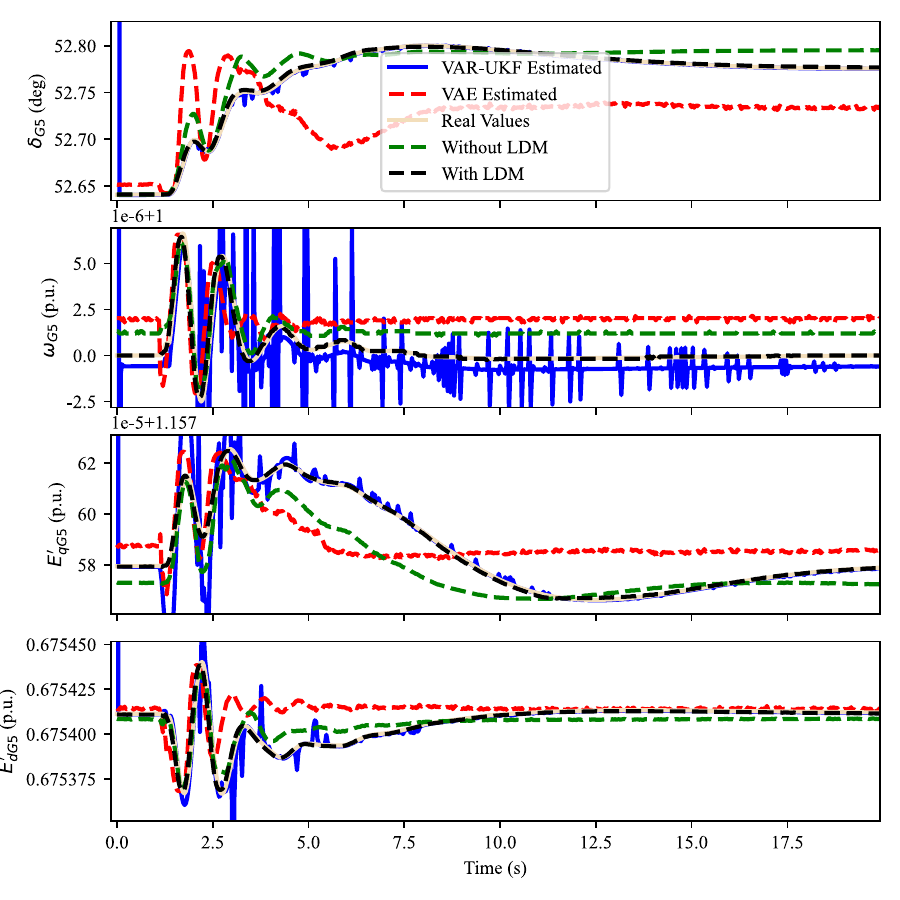}
        \label{res_fig8}
    }
    \vspace{-0.4cm}
    \caption{The estimated results for power angle $\delta$, rotor speed $\omega$, and transient voltage $E'$ in the test system under bad PMU data or communication uncertainty.}
    \vspace{-0.4cm}
\end{figure*}

\subsection{Robustness to Bad PMU Measurements}
Consecutive gross PMU measurement errors or noises can severely affect the accuracy of DSE. In the IEEE 39-bus system, the Line 22 (connecting Buses 16 and 19) was disconnected between 1s and 1.06s. Subsequently, the PMU measurements including voltage magnitude $V$, phasor angle $\theta$, active power $P$, and reactive power $Q$ at terminal Bus 22 of Generator 3 (G3) are subjected to a ramp component with a 2\% peak error between 2s and 4s, while also being impacted by significant measurement noises, with $\bm{R}=0.1\bm{I}$ and $\bm{Q}=1e^{-4}\bm{I}$. As depicted in Fig. \ref{res_fig3}, VAR-UKF exhibits noticeable deviation in the transient voltage estimation, starting with a ramp-like increase at around 2s. The proposed approach without robust encoder $E_{\phi'}(\cdot)$ is also affected. Additionally, a 5\% step component error between 2s and 5s is superimposed on the PMU data for Generator 24 (G24) in the NPCC 140-bus system, located at terminal Bus 23. The system experiences a transient three-phase metal-to-ground fault on Line 26 (connecting Buses 30 and 31) from 1s to 1.09s, after which the protection devices disconnect Line 26 at 1.066s. Afterwards, Line 26 is re-closed at 1.117s. In Fig. \ref{res_fig5}, the state variable fluctuations for G24 are significant. The proposed method without the robust encoder experiences a large phase error due to bad PMU measurements, while the conventional UKF method demonstrates a noticeable step error between 2s and 5s. Once the measurements are restored to normal, the UKF estimate quickly converged to the real values. In summary, the proposed robust encoder approach maintains high estimation accuracy even under conditions of bad PMU measurements.

\subsection{Two-Phase Data Recovery of Latent Space}
The proposed method significantly reduces the communication burden and bandwidth requirements between power plants, substations, estimation centers, and control centers, as it only requires the transmission of a small amount of latent vectors $\textbf{z}$ to perform DSE. However, uncertainties in the transmission process may introduce DSE inaccuracy and unavailability. The utilized LDM approach can enhance the quality of the transmitted at a lower computational complexity compared to TSDM in previous works \cite{10542391}. Assume that the data transmitted from the terminal bus to the estimation center is subject to uncertainties, with 50\% of the data being either corrupted or unreachable. The estimation performances of different approaches under this scenario are depicted in Fig. \ref{res_fig8} and Fig. \ref{res_fig9}.

In Fig. \ref{res_fig8}, a sudden increase of 0.093 p.u. of active load occurs at Bus 8 at 1s, causing a slight dynamic change in the state variables of Generator 5 (G5). In Fig. \ref{res_fig9}, after the disconnection of Lines 108 and 197 in the NPCC system between 1s and 1.04s, they are re-closed (N-2 contingencies). Clearly, when 50\% of the transmitted data is affected by communication uncertainties, the conventional UKF method exhibits significant and random fluctuations due to prediction errors from VAR, although it ultimately converges to the real values. On the other hand, the VAE-based method and DSE without LDM enhancement fail to match the true values due to the poor quality of the latent bottleneck $\textbf{z}$. However, after processing with LDM, the decoded state variable estimates remain robust even under anomalies, accurately reflecting the internal state of the generators in power systems.

\subsection{Generalization for Unforeseen Operation Conditions}
\begin{figure*}[!t]
    \centering
    \subfigure[Estimated states with 50\% of transmitted data contaminated during a two-line outage event in the NPCC 140-bus system.]{
        \includegraphics[width=0.315\textwidth]{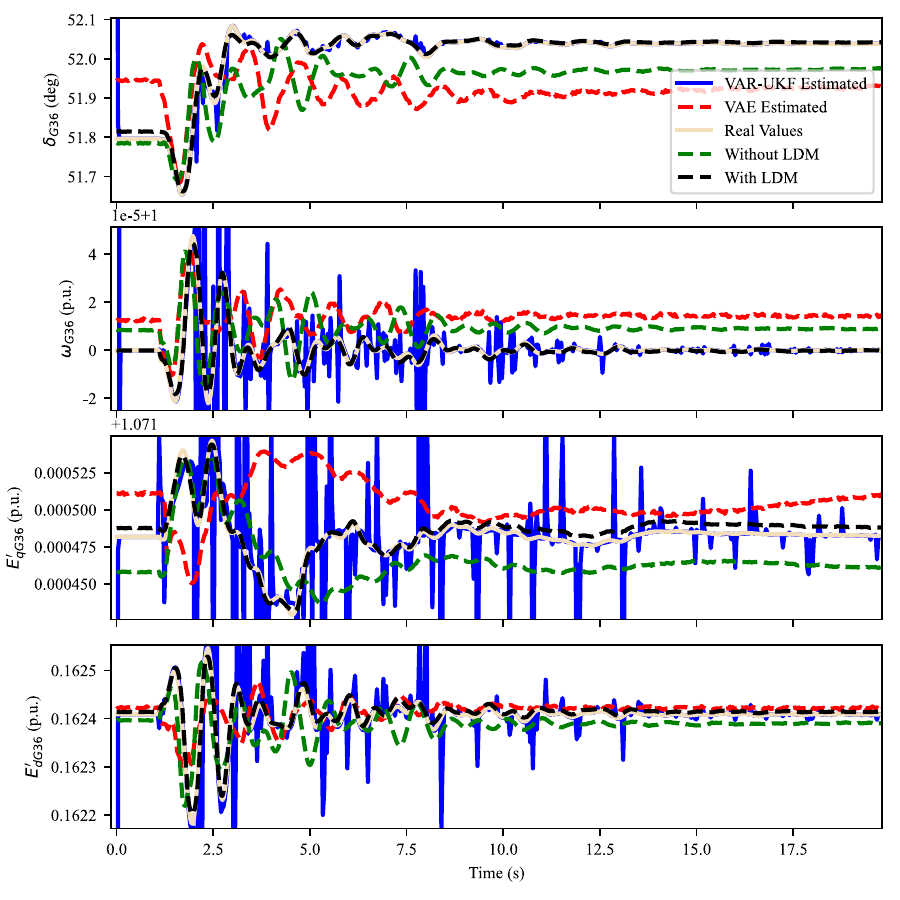}
        \label{res_fig9}
    }
    \subfigure[Estimated states during an unknown generator outage event in the IEEE 39-bus system.]{
        \includegraphics[width=0.315\textwidth]{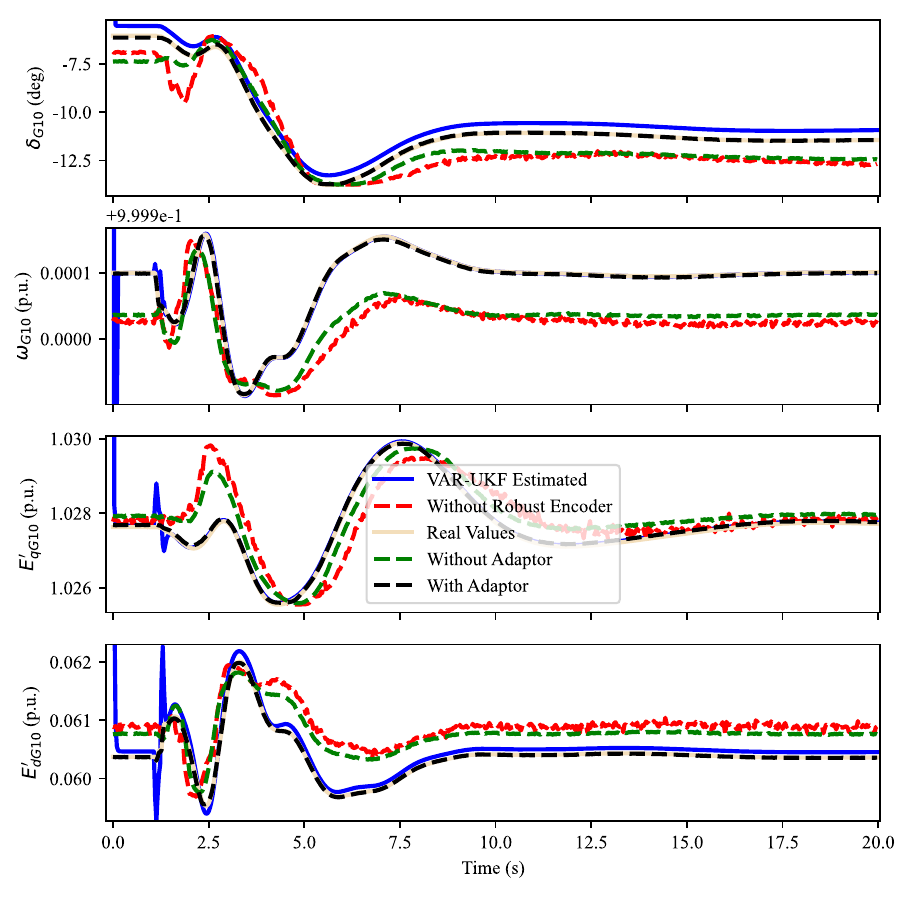}
        \label{res_fig11}
    }
    \subfigure[Estimated states during an unknown generator outage event in the NPCC 140-bus system.]{
        \includegraphics[width=0.315\textwidth]{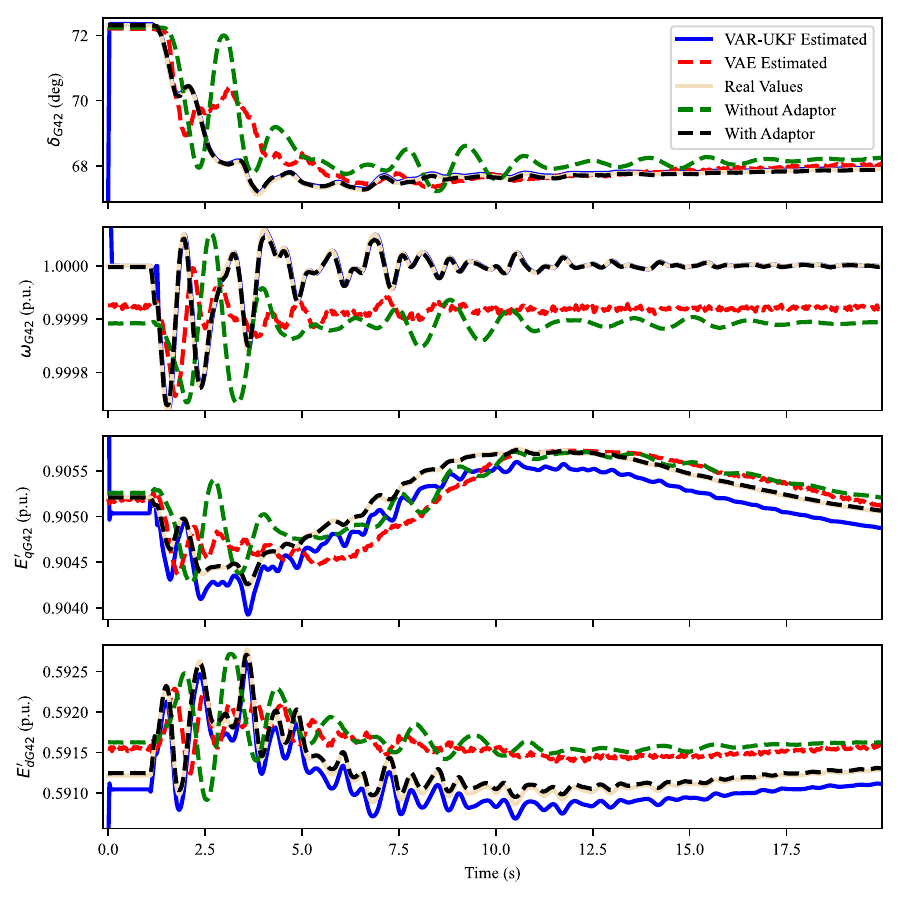}
        \label{res_fig13}
    }
    \vspace{-0.4cm}
    \caption{The estimated results for power angle, rotor speed, and transient voltage in the test system under communication uncertainty and unknown distributions.}
    \vspace{-0.5cm}
\end{figure*}

A major bottleneck of deep learning-aided DSE is its generalization ability to unforeseen power system operating conditions. Specifically, transient events, such as generator outages, are not considered in the model’s training set, making them suitable for testing the effectiveness of the proposed fast adaptation strategy. Traditional DSE approaches are generally more adaptable to various system conditions. However, due to process, topology, or parameter errors arising from changes in system conditions, the larger system process noise covariance matrix $\bm{Q} = 0.1\bm{I}$ is selected during testing to approximate these uncertainties. The results are presented in Fig. \ref{res_fig11} and Fig. \ref{res_fig13}. In Fig. \ref{res_fig11}, the Generator 6 is temporarily disconnected between 1s and 1.09s, showing the transient behavior of Generator 10 (G10), while in the NPCC 140-bus system, Generator 19 is disconnected between 1s and 1.1s. Fig. \ref{res_fig13} illustrates the 20-second state trajectory of Generator 42 (G42).

From the estimation results, it can be observed that the UKF-VAR method, due to the large process noises, shows some deviation from the actual values. Moreover, deep generative models without the adaptor fail to match the true state trajectories when encountering unknown distributions. In contrast, the introduction of the adaptor enables online one-shot learning, allowing $\textbf{z}$ to be fine-tuned based on the current PMU data reconstruction loss. As a result, the proposed method performs exceptionally well under unforeseen power system conditions.

\begin{table}[!h]\centering \footnotesize
	\vspace{-0.3cm}
	\caption{Comparison of the proposed approach and existing methods on computation time (s), error ratio, and data compression ratio over 6000 sampling points.}
	\label{Timeliness}
	\renewcommand{\arraystretch}{0.5}
	\setlength\tabcolsep{0.5em}
	\begin{tabular}{ccccccc}
	\toprule
	\toprule
     \multicolumn{1}{c}{\multirow{2}{*}{Approach}} & \multicolumn{3}{c}{IEEE 39-Bus System} &  \multicolumn{3}{c}{NPCC 140-Bus System}                                               \\
     \cmidrule(lr){2-4}   \cmidrule(lr){5-7} 
     \multicolumn{1}{c}{} & \multicolumn{1}{c}{Time} & \multicolumn{1}{c}{Error} & \multicolumn{1}{c}{Ratio} & \multicolumn{1}{c}{Time} & \multicolumn{1}{c}{Error} & \multicolumn{1}{c}{Ratio} \\
	\midrule
	 UKF-VAR \cite{zhao2019correlation} &  \textbf{4.89} &  9.49\%   & 100\%   &  \textbf{10.35} & 11.47\% & 100\%  \\
	\midrule
	 VAE \cite{khazeiynasab2021power} &  13.06 & 8.33\%   & \textbf{2.67\%}  & 21.79 & 14.72\% & \textbf{1.33\%} \\
	\midrule
	 TSDM \cite{10542391} &  1338 &  1.66\%  & 100\%  & 3126 & \textbf{1.75\%} & 100\%  \\
     \midrule
Proposed Method & \multirow{2}{*}{7.07} & \multirow{2}{*}{8.26\%} & \multirow{2}{*}{\textbf{2.67\%}} & \multirow{2}{*}{12.99} & \multirow{2}{*}{11.35\%} & \multirow{2}{*}{\textbf{1.33\%}} \\
(W/o LDM/Adaptor) \\
    \midrule
    Proposed Method & \multirow{2}{*}{23.93}  &  \multirow{2}{*}{2.47\%}   &  \multirow{2}{*}{\textbf{2.67\%}} & \multirow{2}{*}{31.78} & \multirow{2}{*}{2.19\%} &  \multirow{2}{*}{\textbf{1.33\%}}\\
    (W/ LDM) \\
    \midrule
    Proposed Method & \multirow{2}{*}{38.58}  &   \multirow{2}{*}{\textbf{1.58\%}}  &  \multirow{2}{*}{8.00\%} & \multirow{2}{*}{51.18} & \multirow{2}{*}{1.81\%} & \multirow{2}{*}{4.00\%} \\
    (W/ Adaptor)\\
	\bottomrule
	\bottomrule
	\end{tabular}
	\vspace{-0.3cm}
	\end{table}

\subsection{Compression Ratio and Time Consumption Analysis}
The real-time performance and efficient utilization of communication resources in DSE are critical issues for modern power systems. Traditional centralized DSE requires transmitting all PMU measurements in WAMSs, while deep learning-aided DSE with an encoder-decoder structure only needs to transmit compressed latent bottlenecks $\textbf{z}$. The proposed method improves the accuracy and robustness of DSE under measurement and communication uncertainties by introducing LDM and lightweight adaptors, which inevitably increase computational complexity. However, compared to the TSDM approach, it only requires diffusion in the compressed feature space and simplifies sampling steps through the consistency training, significantly reducing computational overhead. The Dynamic Mean Absolute Percentage Error (DMAPE) denotes the estimation accuracy, which can be written as
\begin{equation} 
	\begin{aligned}
    \textrm{DMAPE} = \frac{1}{K}\sum^K_{k=1}\left |\frac{\hat{\bm{x}}_k-\bm{x}_k}{\bm{x}_{\max}-\bm{x}_{\min}} \right |\times 100\%
	\end{aligned}
\end{equation}
based on the state variations, where $K$ is the total sampling points and $\hat{\bm{x}}_k$ is the estimated states. The performance of the proposed method and other existing approaches, in terms of total computation time, DMAPE, and data compression ratios, is evaluated on the IEEE 39-bus and NPCC 140-bus systems over 20 seconds and 600 sampling points (the sampling frequency $f$=30Hz), as demonstrated in Table \ref{Timeliness}. The results indicate that the proposed approach performs similarly to TSDM, with reduced time consumption and lower DMAPE compared to UKF-VAR, while achieving effective data compression.


\section{Conclusion} 
\label{sec:Conclusion}
Accurate and fast dynamic state estimation (DSE), which relies on PMU data to estimate state variables, is a core element for power system modeling, analysis, monitoring, protection, and control applications. The proposed deep generative model-assisted DSE offers the following improvements: 1) it estimates generator state variables and control input variables jointly by utilizing PMU measurement vector input; 2) it maintains estimation accuracy even under large errors in PMU data; 3) it processes communication anomalies in centralized DSE through the application of latent diffusion model at the estimation center; and 4) it adapts to unforeseen power system events, such as generator-trip, by utilizing a lightweight adaptor to rapidly transform latent features. Extensive experimental results illustrate that the proposed approach effectively addresses the existing DSE challenges, reduces transmitted data redundancy, and further decreases computational complexity compared to the original two-stage diffusion model (TSDM) \cite{10542391}. However, compared to real-time traditional Kalman filtering-based DSE, the proposed model, composed of complex components, inevitably increases the time consumption of DSE. In some scenarios, it may even fail to meet the strict real-time DSE requirements of power systems. To this end, improving the computational efficiency of such complex AIGC models with high accuracy remains an ongoing challenge.


\bibliographystyle{IEEEtran}
\bibliography{IEEEabrv,bibi}

\end{document}